\documentclass[9pt,bestpractices]{livecoms}
\usepackage[utf8]{inputenc}
\usepackage{lipsum} 
\usepackage[version=4]{mhchem}
\usepackage{siunitx}
\DeclareSIUnit\Molar{M}
\usepackage[italic]{mathastext}
\usepackage{amsmath}
\usepackage{amssymb}
\usepackage{subcaption}
\usepackage{pdflscape}
\usepackage[para,flushleft,online]{threeparttable}
\usepackage{colortbl}
\usepackage{xurl}
\graphicspath{{figures/}}

\newcommand{\versionnumber}{0.2}  
\newcommand{\githubrepository}{\url{https://github.com/openforcefield/FE-Benchmarks-Best-Practices}}  
\newcommand\standardstate{{\circ\kern-0.495em-}}

\definecolor{good}{HTML}{6BB11E}
\definecolor{medium}{HTML}{B0E66D}
\definecolor{bad}{HTML}{C44122}

\def\ligcountthres{16}
\def\ligcountthresideal{25}
\def\ligrangethres{3.0}
\def\ligrangethresideal{5.0}

\newcommand{\ligcount}[1]{
    \ifdim #1 pt < \ligcountthres pt%
        {\cellcolor{bad}#1}
    \else
        \ifdim #1 pt < \ligcountthresideal pt%
            {\cellcolor{medium}#1}
        \else
            {\cellcolor{good}#1}
        \fi
    \fi
    }
\newcommand{\ligrange}[1]{
    \ifdim #1 pt < \ligrangethres pt%
        {\cellcolor{bad}#1}
    \else
        \ifdim #1 pt < \ligrangethresideal pt%
            {\cellcolor{medium}#1}
        \else
            {\cellcolor{good}#1}
        \fi
    \fi
    }
    
\newcommand{\notrust}[2]{{\cellcolor{bad} #1 (NT, #2)}}
\newcommand{\mildlytrust}[2]{{\cellcolor{medium} #1 (MT, #2)}}
\newcommand{\hightrust}[2]{{\cellcolor{good} #1 (HT, #2)}}

\newcommand{\rmse}[3]{#1 [#2,#3]}
\title{Best practices for constructing, preparing, and evaluating protein-ligand binding affinity benchmarks [Article v\versionnumber]}
\author[1*]{David F. Hahn}
\author[2]{Christopher I. Bayly}
\author[3,4]{Hannah E. Bruce Macdonald}
\author[3]{John D. Chodera}
\author[5]{Vytautas Gapsys}
\author[6]{Antonia S. J. S. Mey}
\author[7]{David L. Mobley}
\author[1]{Laura Perez Benito}
\author[8]{Christina E. M. Schindler}
\author[1]{Gary Tresadern}
\author[9]{Gregory L. Warren}
\affil[1]{Computational Chemistry, Janssen Research \& Development, Turnhoutseweg 30, Beerse B-2340, Belgium}
\affil[2]{OpenEye Scientific Software, 9 Bisbee Court, Suite D, Santa Fe, NM 87508 USA}
\affil[3]{Computational and Systems Biology Program, Sloan Kettering Institute, Memorial Sloan Kettering Cancer Center, New York, NY 10065 USA}
\affil[4]{MSD, The Francis Crick Institute, 1 Midland Road, London, NW1 1AT, United Kingdom}
\affil[5]{Computational Biomolecular Dynamics Group, Max Planck Institute for Biophysical Chemistry, G\"ottingen, Germany}
\affil[6]{EaStCHEM School of Chemistry, David Brewster Road, Joseph Black Building, The King's Buildings, Edinburgh, EH9 3FJ, UK}
\affil[7]{Departments of Pharmaceutical Sciences and Chemistry, University of California, Irvine, CA USA}
\affil[8]{Computational Chemistry \& Biology, Merck KGaA, Frankfurter Str. 250, 64289 Darmstadt, Germany}
\affil[9]{DeepCure, 131 Dartmouth St, Boston, MA 02116 USA }

\corr{dhahn3@its.jnj.com}{DFH}  

\orcid{David F. Hahn}{0000-0003-2830-6880}
\orcid{Christopher I. Bayly}{0000-0001-9145-6457}
\orcid{Hannah E. Bruce Macdonald}{0000-0002-5562-6866}
\orcid{John D. Chodera}{0000-0003-0542-119X}
\orcid{Vytautas Gapsys}{0000-0002-6761-7780}
\orcid{Antonia S. J. S. Mey}{0000-0001-7512-5252}
\orcid{David L. Mobley}{0000-0002-1083-5533}
\orcid{Laura Perez Benito}{0000-0001-9607-9048}
\orcid{Christina E. M. Schindler}{0000-0002-8980-048X}
\orcid{Gary Tresadern}{0000-0002-4801-1644}
\orcid{Gregory L. Warren}{0000-0003-4017-0162}

\blurb{This LiveCoMS document is maintained online on GitHub at \githubrepository; to provide feedback, suggestions, or help improve it, please visit the GitHub repository and participate via the issue tracker.}

\pubDOI{10.XXXX/YYYYYYY}
\pubvolume{<volume>}
\pubissue{<issue>}
\pubyear{<year>}
\articlenum{<number>}
\datereceived{Day Month Year}
\dateaccepted{Day Month Year}

\begin{document}

\begin{frontmatter}
\maketitle

\begin{abstract}
Free energy calculations are rapidly becoming indispensable in structure-enabled drug discovery programs. 
As new methods, force fields, and implementations are developed, assessing their expected accuracy on real-world systems (\emph{benchmarking}) becomes critical to provide users with an assessment of the accuracy expected when these methods are applied within their domain of applicability, and developers with a way to assess the expected impact of new methodologies.
These assessments require construction of a benchmark---a set of well-prepared, high quality systems with corresponding experimental measurements designed to ensure the resulting calculations provide a realistic assessment of expected performance when these methods are deployed within their domains of applicability. 
To date, the community has not yet adopted a common standardized benchmark, and existing benchmark reports suffer from a myriad of issues, including poor data quality, limited statistical power, and statistically deficient analyses, all of which can conspire to produce benchmarks that are poorly predictive of real-world performance.
Here, we address these issues by presenting guidelines for (1) curating experimental data to develop meaningful benchmark sets, (2) preparing benchmark inputs according to best practices to facilitate widespread adoption, and (3) analysis of the resulting predictions to enable statistically meaningful comparisons among methods and force fields.
We highlight challenges and open questions that remain to be solved in these areas, as well as recommendations for the collection of new datasets that might optimally serve to measure progress as methods become systematically more reliable.
Finally, we provide a curated, versioned, open, standardized benchmark set adherent to these standards ({\bf protein-ligand-benchmark}) and an open source toolkit for implementing standardized best practices assessments ({\bf openff-arsenic}) for the community to use as a standardized assessment tool.
While our main focus is free energy methods based on molecular simulations, these guidelines should prove useful for assessment of the rapidly growing field of machine learning methods for affinity prediction as well.
\end{abstract}

\end{frontmatter}


\section{Overview}

\begin{figure*}[!ht]
    \centering
    \includegraphics[width=0.95\linewidth]{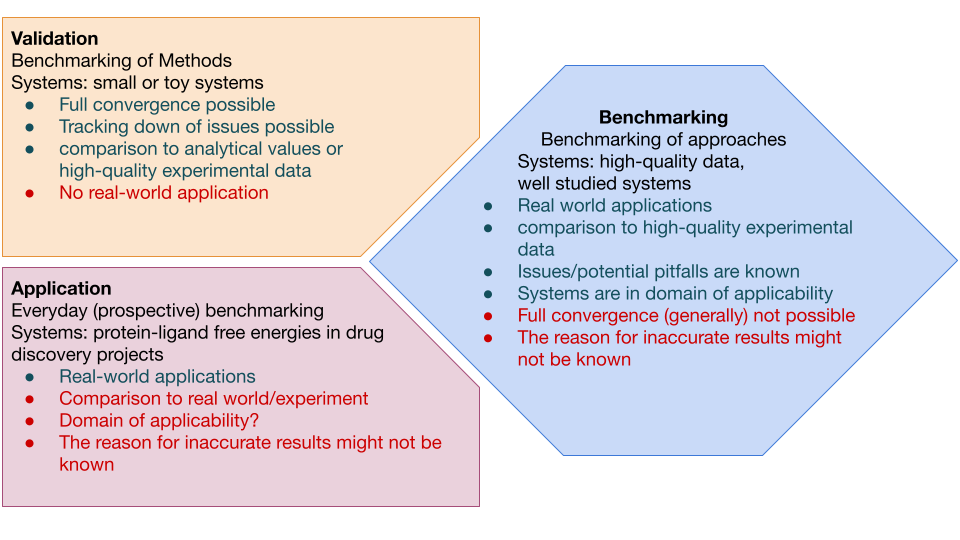}
    \caption{
    \textbf{Illustration of the definitions of \textit{Validation}, \textit{Application}, and \textit{Benchmarking} used in this guide.} 
    For each term, the definition, advantages (green) and potential short-comings (red) in terms of method evaluation are listed in the three panels.
    \textit{Validation} (top left panel) uses systems that will confidently converge, the expected results are known, and the underlying issues are well understood. 
    Validation sets allows robust development and improvement of methods.
    \textit{Application} (bottom left panel) of a method, on the other hand, uses real-world systems and enables methods to be continuously evaluated on real-world applications of interest.
    Because the systems may not be well understood, it is possible for methods to fail in new ways that are difficult to detect.
    \textit{Benchmarking} (right panel) bridges validation and application by aiming to assess the accuracy of real-world applications relative to experiment in cases where experimental data quality is not limiting and the method is known to be applied within its domain of applicability. 
    Compared to validation, the size and complexity of the system may introduce challenges to producing robust, repeatable results.
    }%
    \label{fig:benchmarking_definition}
\end{figure*}

This guide focuses on recommended best practices for benchmarking the accuracy of small molecule binding free energy (FE) calculations. 
Here, we define \emph{benchmarking} as the assessment of expected real-world performance relative to experiment.
We contrast this with the assessment of methods or tools intended to arrive at the same target free energy, which we refer to as \emph{validation} (Figure~\ref{fig:benchmarking_definition}), the comparison of the computational efficiency or speed of these methods, or mapping of effort-accuracy trade offs, all of which also play essential roles in dictating real-world usage.
Importantly, validation calculations are often performed on systems selected for tractability, rather than intended to be representative of real-world applications~\cite{mobleyPredictingBindingFree2017,vangunsteren_validation_2018,tsai_validation_2020}.

%

As illustrated in Figure~\ref{fig:benchmarking_definition}, benchmarking against experiment would ideally be performed on high quality data in order to provide an accurate assessment of expected performance under conditions where structure or assay deficiencies do not limit performance. 
In good benchmark sets, the potential pitfalls and complications in the data are well understood, but these systems may still challenge methodologies to produce reproducible, consistent predictions due to conformational sampling timescales---unlike simpler systems selected for methodology validation.
We also differentiate benchmarking from \emph{application} (Figure~\ref{fig:benchmarking_definition}), where one is often constrained by the availability of experimental data and limited to a particular target, which may not always fall within the domain of applicability of the methodology. 
We aim to construct benchmarks that provide a good predictor of the expected accuracy in applications that fall squarely within the domain of applicability and for which good experimental data is available.

\underline{Organization:} 
This best practices guide is organized as follows:
First, we give a brief overview of protein-ligand binding free energy methods and their use with the goal of highlighting key concepts that guide the construction of a meaningful benchmark.
Next, we discuss recommendations for the construction of a high-quality experimental benchmark dataset, which must consider the availability of high-quality structural and bioactivity data as well as the expected domain of applicability.
Next we provide recommendations on preparing structures for free energy calculations in a manner that will enable the benchmark dataset to be widely and readily usable by practitioners and developers, incorporating best practices for carrying out free energy calculations.
We then discuss recommendations for the statistical analysis of both retrospective benchmarks and blind prospective challenges in order to derive robust conclusions about the accuracy of these methods and insights into where they fail.
To address the absence of a standard community-wide benchmark, we provide a curated, versioned, open, standardized benchmark set adherent to these standards ({\bf protein-ligand-benchmark}). In addition, we provide an open source toolkit that implements standardized best practices for assessment and analysis of free energy calculations ({\bf openff-arsenic}).
Finally, we conclude with recommendations for data collection and curation to guide the systematic improvement of available benchmark sets and  drive the expansion of the domain of applicability of free energy methods.

\section{Introduction}
\label{sec:intro}

The quantitative prediction of protein ligand binding affinity is a key task in computer-aided drug discovery (CADD). 
Accurate predictions of ligand affinity can significantly accelerate early stages of drug discovery programs when used to prioritize compounds for synthesis with the goal of improving or maintaining potency~\cite{abelCriticalReviewValidation2017,abelModelingValuePredictive2018}. 
Binding free energy calculations---particularly alchemical binding free energy calculations---have emerged as arguably the most promising tool~\cite{courniaRelativeBindingFree2017}. 
Alchemical methods, which include a multitude of approaches such as free energy perturbation (FEP)~\cite{zwanzigHighTemperatureEquation1954,bennettEfficientEstimationFree1976} and thermodynamic integration (TI)~\cite{kirkwoodQuantumStatisticsAlmost1933,kirkwoodQuantumStatisticsAlmost1934,kirkwoodStatisticalMechanicsFluid1935}, have a substantial legacy, with the original theory dating back many decades. 
Seminal work in the 1980’s and 90’s demonstrated that molecular dynamics (MD) and Monte Carlo (MC) simulation packages could carry out these calculations for practical applications in organic and biomolecular systems~\cite{jorgensenMonteCarloSimulation1985,straatsmaFreeEnergyHydrophobic1986,lybrandTheoreticalCalculationRelative1986,merzFreeEnergyPerturbation1989,pearlmanDeterminationDifferentialEffects1995,choderaAlchemicalFreeEnergy2011,mobleyPerspectiveAlchemicalFree2012}. 

Alchemical perturbations in binding free energy calculations involve the transformation of one chemical species into another, or its complete creation or deletion, via a chemically unrealistic pathway (\emph{alchemical}) that can only be achieved \textit{in silico} by manipulating its interactions in a defined way. This is achieved by changing an atom from one element identity to another.
Alchemical calculations are often classified as either \emph{relative} (RBFE) or \emph{absolute} (ABFE) binding free energy calculations. 
While the underlying theory is similar, the implementation differs in how the thermodynamic cycle is constructed and which quantities can be computed:
In relative calculations (RBFEs), a generally modest alchemical transformation of the chemical substructures that differ between the ligands is performed to compute the difference in free energy of binding between the related ligands ($\Delta \Delta G$).
By contrast, absolute calculations (ABFEs) alchemically remove an entire ligand, enabling the absolute binding free energy of a ligand ($\Delta G$) to be computed and directly compared to experiment.
A detailed review of commonly-used alchemical methodologies and best practices for their use is provided in a separate best practices guide~\cite{meyBestPracticesAlchemical2020}.

In drug discovery, \emph{lead optimization} (LO) typically involves the synthesis of hundreds of close analogues, often differing by only small structural modifications, in order to identify the optimal leads that show a good balance of target potency and other properties. 
This makes it an ideal scenario for RBFE, where small differences in structure are well suited to alchemical perturbation. 

A number of recent studies have highlighted the good performance of RBFE for LO tasks. 
An early influential publication from Schr\"{o}dinger~\cite{wangAccurateReliablePrediction2015} reported mean unsigned errors of < 1.2 kcal/mol on a curated set of 8 protein targets, 199 ligands, and 330 perturbations using their commercial implementation of FEP.
Minimal discussion was devoted to \emph{how} these targets were selected, other than their diversity and the availability of published structural and bioactivity data for a congeneric series for each target; notably, some ligands appearing in the published studies from which the data were curated were omitted due to the presence of presumed changes in net charge and the potential for multiple binding modes that would fall outside the domain of applicability.
Schr\"{o}dinger utilized the same benchmark set to assess subsequent commercial force field releases (OPLS3~\cite{harder_opls3_2016} and OPLS3e~\cite{roos_opls3e_2019}). 
In the absence of other significant efforts to curate benchmark sets, this set (often called the "Schr\"{o}dinger JACS set") has become the \emph{de facto} dataset for most large scale RBFE reports, used to compare the performance of Amber/TI calculations~\cite{songUsingAMBER18Relative2019}, Flare’s FEP (a collaboration between Cresset and the Michel group)~\cite{kuhnAssessmentBindingAffinity2020}, and PMX/Gromacs~\cite{gapsysLargeScaleRelative2020}, as well as machine learning studies~\cite{jimenezDEEPProteinLigand2018,jimenez-lunaDeltaDeltaNeuralNetworks2019}. 
By contrast, ABFE calculations have not been studied on datasets of similar scale to date, although individual reports have shown success accurately predicting binding affinities~\cite{aldeghiLargescaleAnalysisWater2018,courniaRigorousFreeEnergy2020}.

Despite the reported success of RBFE calculations on these benchmark sets, there are many reports demonstrating that RBFE calculations still struggle in scenarios~\cite{sherborne_collaborating_2016} such as with scaffold modifications~\cite{wangAccurateModelingScaffold2017}, ring expansion~\cite{liuRingBreakingFeasible2015}, water displacement~\cite{michel_energetics_2009,brucemacdonald_ligand_2018,ross_enhancing_2020a,ben-shalom_accounting_2020}, protein flexibility~\cite{huang_insights_2012,fratev_improved_2019,singh_absolute_2020}, applications to GPCRs~\cite{lenselink_predicting_2016,deflorian_accurate_2020}, and the modelling of cofactors such as metal ions or heme~\cite{swiderek_binding_2011,ono_improvement_2020}. 
This is manifested in a large-scale study of FEP applied to active drug discovery projects at Merck KGaA, in which Schindler et al.\ reported several cases of disappointing outcomes~\cite{schindler_largescale_2020}. 

In addition, new methods and implementation improvements for FE calculations continue to emerge, for instance the efforts on lambda dynamics~\cite{knightMultisiteDynamicsSimulated2011,vilseckPredictingBindingFree2018}, and non-equilibrium RBFE calculations~\cite{gapsysLargeScaleRelative2020,rufaChemicalAccuracyAlchemical2020}. 
Furthermore, there are many other methodologies such as end-point binding FE calculations (for instance MMGBSA, MMPBSA) or pathway based FE calculations that continue to be developed and applied.~\cite{genheden_MM_2015} Therefore, we must balance the increased confidence that simulation-based FE calculations can impact drug discovery, with the need to further understand, test, and overcome limitations of the current methods.

In brief, the issues mentioned above are related to three challenges for FE calculations,
(1) an accurate representation of the biological system, 
(2) an accurate force field, and 
(3) sufficient sampling. 
Therefore, despite the importance of FE methods to drug discovery and chemical biology it is surprising that there are no benchmark sets or standard benchmark methodologies that allow calculation approaches to be compared in a manner that will reflect their future performance. 

The Drug Design Data Resource~\cite{amaro_drug_2021} (D3R) and Statistical Assessment of the Modeling of Proteins and Ligands~\cite{mobley_sampl_2021} (SAMPL) prospective challenges have demonstrated the utility of focusing the community on common benchmark systems and using common methods to analyze performance~\cite{geballe_sampl2_2010,muddana_blind_2012,muddana_prediction_2012,muddana_sampl4_2014,gathiaka_d3r_2016,bannan_blind_2016,yin_overview_2017,gaieb_d3r_2018,gaieb_d3r_2019,parks_d3r_2020}. Mobley and Gilson discussed the need for well-chosen validation datasets and how this will have multiple benefits to understanding and expanding the domain of applicability of FE methods~\cite{mobleyPredictingBindingFree2017}. They focused on validation systems that will confidently converge, and where the underlying issues are well understood. The aim was to describe systems that could be used only to assess method performance in a robust manner. As mentioned above, here we define benchmarking as assessing accuracy relative to experiment. This has implications that will be discussed in more detail throughout this article, for instance, the reliability of the underlying experimental data (structure and bioactivities), the confidence in the system setup such as protein and ligand preparation, the suitability of alchemical perturbations for FE, the statistical power of the dataset, the ability of the datasets to capture challenging real-world applications, and recommendations for analysing results. Essentially, we seek to understand what performance can be achieved when all these variables are handled to the best of our abilities.    

Here, our proposed benchmark set augments existing datasets while recommending cleaning up or removing entirely some protein-ligand sets. We highlight key considerations in the construction of a useful set of protein-ligand benchmarks and the preparation of these systems for use as a community-wide benchmark. These recommendations are mirrored in a living benchmark set, which can be used to reliably launch future studies~\cite{dfhahn_openforcefield_2021}. We seek to improve the initial version of this benchmark set in the future with help of the whole community. We welcome any contribution either to improve the existing set or to expand the set with new protein-ligand sets, if they meet the requirements established in here. 
We also recommend statistical analyses for assessing and comparing the accuracy of different methods and provide a set of open source tools that implement our recommendations~\cite{github_openforcefield_arsenic_2020}. 
We hope these materials will become a common standard utilized by the community for assessing performance and comparing methodologies.

\section{Prerequisites}
We assume a basic familiarity with molecular dynamics (MD) simulations, as well as alchemical free energy protocols. 
If you are unfamiliar with both of these concepts we suggest the best practices guides by Braun et al.~\cite{braunBestPracticesFoundations2019} on molecular simulations and Mey et al.~\cite{meyBestPracticesAlchemical2020} on alchemical free energy calculations as a starting point.

\section{Dataset selection}
\label{sec:dataset}

Details of our criteria for the construction of good benchmark datasets will follow throughout the rest of the manuscript. 
Here, we examine the purpose of protein-ligand benchmark datasets, and the rationale for expanding these sets.
We propose a core of robust datasets that match our suggested optimal criteria for benchmarking, but emphasize the need to supplement this core with new datasets which explore increasingly difficult challenges in order to continue to expand the domain of applicability of predictive methods. 
A variety of parameters can guide future datasets.

\subsection{Protein Selection}
\label{sec:dataset:proteins}
The selection of target proteins in the benchmark set is generally dependent
on the availability of experimental data and whether the applied methods are applicable to the specific targets.
A good benchmark system (consisting of a protein target and small molecules with available experimental binding data) should ideally be representative of classical drug discovery targets and chemistry; a good benchmark set should also be diverse in terms of targets and chemistry. 
Expansion of this set to include additional systems should ideally reflect the evolution of drug discovery and the emergence of new target families and chemistries.
%
While binding free energy calculations are agnostic to protein classification, there can be a pragmatic value in expanding benchmark sets to new protein families that may present unexpected inherent difficulties (see Section~\ref{sec:dataset:challenges}).

To merit inclusion in a good benchmark set, the available structural data must meet certain quality thresholds (Section~\ref{sec:struct_data}), and the structure should be adequately prepared for molecular simulation to enable the benchmark to be broadly and readily useful (Section~\ref{sec:prep}).

\subsection{Ligand Selection}
\label{sec:dataset:ligands}

\begin{figure}[!ht]
    \centering
    \includegraphics[width=.48\textwidth]{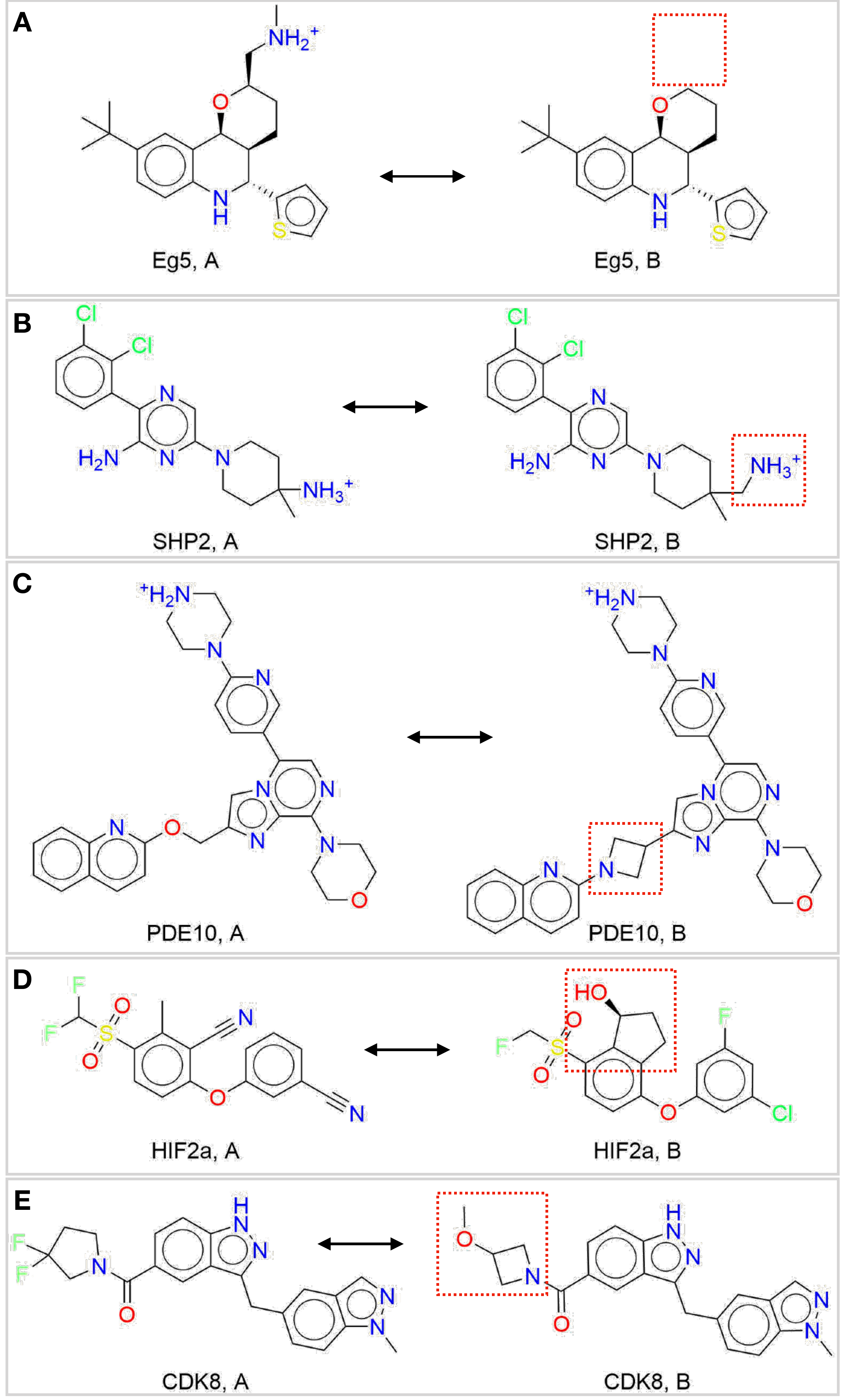}
    \caption{\textbf{Five ligand pairs (A, B) for different targets (with each pair for a single target)
    having structural differences which can be challenging to simulate.}
    (\textbf{A}) Eg5: charge change,
    (\textbf{B}) SHP2: charge move,
    (\textbf{C}) PDE10: linker change,
    (\textbf{D}) HIF2$\alpha$: ring creation,
    (\textbf{E}) CDK8: ring size change. 
    }
    \label{fig:difficult_perturbations}
\end{figure}

While some methods (such as machine learning and GBSA rescoring) can make rapid predictions of affinity, free energy methods are generally relatively costly in terms of computational effort. 
In order to make statistically meaningful comparisons among methods, however, a sufficient number of reliable experimental measurements (Section~\ref{sec:affinities}) will be necessary for a benchmark set. These measurements also need to cover an adequate dynamic range, i.e. the activity range should be sufficiently large.   Such a set enables a statistical analysis with sufficient power to distinguish how methods are expected to perform on larger test sets for the same targets (Section~\ref{sec:analysis}).
In addition, the set of ligands should be both unambiguously specified (with resolved stereochemistry or ambiguous tautomeric or protonation states) and have chemistries that fall within the domain of applicability of the particular free energy method used. 
In order for standardized benchmark sets to be broadly applicable to a range of methodologies and software packages, we recommend annotating systems in terms of common challenges that may exclude their assessment by certain methods or packages.
For relative free energy calculations, these labels should denote transformations that include
(1) charge changes, 
(2) change of the location of a charge,
(3) ring breaking,
(4) changes in ring size, 
(5) linker modifications,
(6) change in binding mode, 
and (7) irreversible (covalent) inhibitors. 
Several of these issues are illustrated in Figure~\ref{fig:difficult_perturbations}.
If the ligand sets are sufficiently large, they can then be split into separate subsets (subsets with e.g., different ring sizes or different charges).

Adequately sampling ligand conformers can pose a challenge for some methods, especially if the ligands contain many rotatable bonds, invertible stereocenters, or macrocycles.
Aromatic rings with asymmetric substitution will usually sample dihedral rotations freely in solvent, but in complex can become trapped in protein pockets during short simulations~\cite{kaus_how_2015,sasmal_sampling_2020}. 
Barriers to inversion of pyramidal centers can sometimes be long compared to typical simulation timescales~\cite{koeppl1967inversion}.
Macrocycles present more extreme challenges for ligand sampling, and likely require special consideration to ensure their conformation spaces are adequately sampled~\cite{wagner_computational_2017,yu_accurate_2017,paulsen_evaluation_2020}. 

The chemical diversity of ligands considered for inclusion in a benchmark set also needs to be suitable for the given free energy method. 
Single RBFEs rely on common structural elements between the molecules being compared, and are hence more appropriate for a congeneric series of ligands. 
ABFEs are more amenable for comparing sets of small molecules that differ more substantially in scaffold, or where the common structural elements are minimal. 
In both kinds of calculations, the size of the structural elements that differ between ligands within a congeneric series is also important to consider, since larger changes may also affect the binding mode of the ligand; the quality and availability of crystal structures for representative ligands of this system becomes critical in assessing these assumptions.

\subsection{Addressing specific challenges}
\label{sec:dataset:challenges}

Besides the challenges mentioned in Sections \ref{sec:dataset:proteins} and \ref{sec:dataset:ligands},
there are specific challenges which can be addressed by a benchmark set. These include 
water displacement in binding sites,
the presence of cofactors in the binding site, 
slow motions of ligands (e.g. rotatable bonds) and proteins, and 
activity cliffs.
We recommend annotating these challenging cases in the benchmark set.




\subsection{Structural Data}
\label{sec:struct_data}

A successful free energy calculation requires a well-prepared, experimentally accurate model of the system to be simulated, with structure(s) representative of the equilibrium state of the system. 
Just as choices made selecting binding data are critical, the choices made when selecting a protein model will impact benchmarking.

Often structural studies use shorter constructs that might be missing several domains compared to the full-length protein. To facilitate crystallization or expression, mutations might have been introduced. In addition, parts of the protein might not be resolved or modelled in available structures. Ideally, such deviations should be kept to a minimum in a benchmark dataset.

\begin{figure}[!ht]
    \centering
    \includegraphics[width=.47\textwidth]{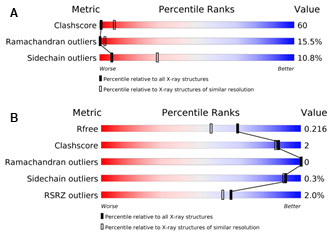}
    {
        \phantomsubcaption\label{fig:jnk1_pdb_report_2gmx}
        \phantomsubcaption\label{fig:jnk1_pdb_report_3eljx}
    }
    
    \caption{\textbf{The PDB structure validation report percentile score panels for the Jnk1 structures PDB IDs 2GMX and 3ELJ from the RCSB PDB.} \textbf{(\subref{fig:jnk1_pdb_report_2gmx})} Note that 2GMX is a poorly ranked structure relative to all structures of similar resolution in the PDB.  
    \textbf{(\subref{fig:jnk1_pdb_report_3eljx})} In contrast 3ELJ is as good or better than structures of similar resolution or all structures in the PDB.}
    \label{fig:jnk1_pdb_report}
\end{figure}

Starting structures are typically obtained from experimentally constrained models, most commonly from X-ray diffraction data.
Other sources include cryo-EM, NMR or homology models~\cite{courniaRelativeBindingFree2017,courniaRigorousFreeEnergy2020,schindler_largescale_2020}.
As free energy calculations are usually run at atomic resolution, the input structure needs to provide the coordinates of all atoms, with those coordinates ideally determined by the experimental model.
For X-ray and cryo-EM structures, this requirement is only met by high quality structures.
The evaluation criteria defined by OpenEye Iridium~\cite{warrenEssentialConsiderationsUsing2012} can guide the assessment of X-ray structures. The lower the Iridium score, the better the quality of the structure. The Iridium classification categorizes each structure into not trustworthy (NT),
mildly trustworthy (MT) and highly trustworthy (HT) categories. It is important to note that the Iridium criteria were designed to assess structures for benchmarking docking and not necessarily for free energy calculations. As such there is one important criterion missing -- completeness of the model -- which is likely to be far more important for free energy calculations than docking.

\begin{table*}[!ht]
\centering
\caption{
\textbf{Evaluation of the quality of structural and activity experimental data of the proposed benchmark set.}
The structures listed in "Used structure" are those used in the initial version of this dataset, which is drawn in part from previous studies. However, alternate available structures may be superior. In these cases, we provide a PDB ID of a higher quality structure and 
its quality measures, in the "Alternate structure" field.
The alternate structures are sorted according to best structure (lowest Iridium score) first. The footnotes "b" denote structures with similar ligands as the used structure. 
For each structure, the PDB ID is followed with the 
Iridium classification and Iridium score in the brackets.
The Iridium classification categorizes each structure into not trustworthy (NT),
mildly trustworthy (MT) and highly trustworthy (HT) categories. The lower the Iridium score, the better the structure~\cite{warrenEssentialConsiderationsUsing2012}.
For the used structures, also the 
diffraction-component precision indeces (DPI) is listed.
We define a high ligand similarity as the OpenEye TanimotoCombo (Shape and Color Tanimoto, range from 0 to 2) being larger than 1.4 (standard cutoff).
Regarding activity data ("Ligand Information"), the following metrics are given:
The number of ligands N, 
the dynamic range DR ($\mathrm{DR}=\mathrm{max}(\Delta G)-\mathrm{min}(\Delta G)$),
and a simulated RMSE .
For the calculation of the simulated RMSE, predicted $\Delta G$ data was drawn from a Gaussian distribution around the experimental value with a standard deviation of $\sigma = 1\,\mathrm{kcal\,mol^{-1}}$, taking also the experimental error into account. The numbers in the brackets are the 95\% confidence intervals, obtained by bootstrapping using 1000 bootstrap samples. 
The quality metrics are color coded to highlight ideal quality (dark green), minimum quality (light green) and low quality (red). The ideal and minimum quality codes correspond to the minimal and ideal requirements of the checklist "Minimal requirements for a dataset".
}
\begin{threeparttable}
\def\present{a}
\def\similar{b}
\def\lesssimilar{c}
\def\crystal{d}
\def\packing{e}
\def\jnkfig{f}
\def\alternate{g}
\def\ligdens{h}
{\footnotesize
\begin{tabular}[t]{l|ll|ll|lll}
Target     & \multicolumn{2}{c|}{Used structure}                          & \multicolumn{2}{c|}{Alternate structures} & \multicolumn{3}{c}{Ligand Information}\\
           &PDB & DPI & 
           &  
           & N 
           & DR  
           & RMSE\\
           & &  &               & & & \multicolumn{2}{l}{[$\mathrm{kcal\,mol^{-1}}$]} \\
\hline
BACE\cite{cumming_structure_2012,wang_accurate_2015}
    &\hightrust{4DJW}{0.32}   
    & 0.11 
    &  \begin{tabular}[t]{p{9em}} 
         \hightrust{ 6UWP}{0.28} \\
         \hightrust{3TPP}{0.28}\tnote{\present} \\
         \hightrust{4DJV}{0.31}\tnote{\similar} \\ \hightrust{3INH}{0.33}\tnote{\similar} 
        \end{tabular}
    & \begin{tabular}[t]{p{9em}} 
         \hightrust{3INF}{0.33}\tnote{\similar} \\
         \hightrust{3IN3}{0.36}\tnote{\present,\similar} \\ 
        \hightrust{3LHG}{0.36}\tnote{\similar}
      \end{tabular} 
    & \ligcount{36}  
    & \ligrange{4.0}
    & \rmse{0.98}{0.78}{1.18} 
    \\
%
BACE\_HUNT\cite{hunt_spirocyclic_2013,ciordia_application_2016,keranen_acylguanidine_2017} &        \hightrust{4JPC}{0.32}          & 0.12 
    & \begin{tabular}[t]{p{9em}}
        \hightrust{6UWP}{0.28} \\
        \hightrust{3TPP}{0.28}\tnote{\present} \\ 
        \hightrust{4JP9}{0.31}\tnote{\present,\similar}
      \end{tabular}
    & \begin{tabular}[t]{p{9em}} 
        \hightrust{4JOO}{0.33}\tnote{\similar} \\
        \hightrust{4RRO}{0.33}\tnote{\similar} \\
        \hightrust{4JPE}{0.35}\tnote{\similar}
      \end{tabular}  
    & \ligcount{32} 
    & \ligrange{4.9}
    & \rmse{0.97}{0.73}{1.19}
    \\
BACE\_P2\cite{malamas_design_2010,keranen_acylguanidine_2017}    
    &\hightrust{3IN4}{0.59}        
    & 0.28 
    & \begin{tabular}[t]{p{9em}} 
        \hightrust{6UWP}{0.28} \\
        \hightrust{3TPP}{0.28}\tnote{\present} \\
        \hightrust{ 4DJV}{0.31}\tnote{\similar}
      \end{tabular}
    & \begin{tabular}[t]{p{9em}}
         \hightrust{3INF}{0.33}\tnote{\similar} \\
         \hightrust{3IN3}{0.36}\tnote{\present,\similar} \\ 
        \hightrust{3LHG}{0.36}\tnote{\similar}
      \end{tabular} 
    & \ligcount{12}  
    & \ligrange{0.8}
    & \rmse{1.00}{0.59}{1.33}
    \\
CDK2\cite{hardcastle_substituted_2004,wang_accurate_2015}       &       \mildlytrust{1H1Q}{0.87}               & 0.28 
    & \begin{tabular}[t]{p{9em}} 
        \hightrust{3DDQ}{0.31}\tnote{\present}
      \end{tabular}
    & \begin{tabular}[t]{p{9em}} 
        \hightrust{4EOR}{0.39}\tnote{\present,\similar}
      \end{tabular} 
    & \ligcount{16}  
    & \ligrange{4.3}
    & \rmse{1.00}{0.67}{1.29}
    \\
CDK8\cite{schiemann_discovery_2016,schindler_largescale_2020}       & \mildlytrust{5HNB }{0.74}          & 0.22 
    & \begin{tabular}[t]{p{9em}} 
        \hightrust{5XS2}{0.33} \\
        \hightrust{5IDN}{0.36}\tnote{\lesssimilar}
      \end{tabular}
    & \begin{tabular}[t]{p{9em}} 
        \hightrust{4CRL}{0.42}\tnote{\present} \\
      \end{tabular} 
    & \ligcount{33}  
    & \ligrange{5.7} 
    & \rmse{0.96}{0.73}{1.18}
\\
c-MET\cite{schindler_largescale_2020}                               & \mildlytrust{4R1Y}{0.75}\tnote{\ligdens}          & 0.17
    & \begin{tabular}[t]{p{9em}} 
        \hightrust{5EOB}{0.28} \\
        \hightrust{3I5N}{0.37}\tnote{\present,\lesssimilar} \\
        \hightrust{3ZC5}{0.38}\tnote{\present,\lesssimilar} \\
        \hightrust{3CD8}{0.39}\tnote{\present,\lesssimilar} \\
      \end{tabular}
    & \begin{tabular}[t]{p{9em}} 
        \hightrust{3ZXZ}{0.44}\tnote{\present,\lesssimilar} \\
        \hightrust{4DEG}{0.46}\tnote{\present,\lesssimilar} \\
        \hightrust{5EYD}{0.47}\tnote{\present,\lesssimilar} \\
        \end{tabular} 
    & \ligcount{24}  
    & \ligrange{6.2} 
    & \rmse{0.99}{0.72}{1.26}
\\
EG5\cite{schiemann_discovery_2010,schindler_largescale_2020}        & \mildlytrust{3L9H}{0.88}          & 0.18 
    & \begin{tabular}[t]{p{9em}} 
        \hightrust{2X7C}{0.32} \\
        \hightrust{3K5E}{0.35}\tnote{\present} 
      \end{tabular}
    & \begin{tabular}[t]{p{9em}} 
        \hightrust{3K3B}{0.41}\tnote{\lesssimilar}
      \end{tabular} 
    & \ligcount{28} 
    & \ligrange{3.5}
    & \rmse{0.98}{0.72}{1.22}\\
Galectin\cite{delaine_galectin3binding_2016,manzoni_assessing_2018}   & \mildlytrust{5E89}{1.04}          & 0.07 
    & \begin{tabular}[t]{p{9em}} 
        \hightrust{5NF7}{0.30} \\
        \hightrust{1KJR}{0.30}\tnote{\present} \\
        \mildlytrust{5ODY}{0.33}\tnote{\similar, \packing, \alternate}
      \end{tabular}
    & \begin{tabular}[t]{p{9em}} 
       \mildlytrust{4BM8}{0.38}\tnote{\present, \crystal,\similar} \\
       \hightrust{ 5OAX}{0.54}\tnote{\similar} \\
      \end{tabular}	
    & \ligcount{8}  
    & \ligrange{2.7}
    & \rmse{1.04}{0.55}{1.42}
    \\ 
HIF2a\cite{wallace_smallmolecule_2016,schindler_largescale_2020}      
    & \hightrust{5TBM}{0.35} 
    & 0.17 
    & \begin{tabular}[t]{p{9em}} 
        \hightrust{3H82}{0.30}\tnote{\present} \\
        \hightrust{6D09}{0.35}\tnote{\similar}
      \end{tabular}
    & \begin{tabular}[t]{p{9em}} 
        \hightrust{5UFB}{0.36}\tnote{\similar}\\
      \end{tabular} 
    & \ligcount{42}  
    & \ligrange{4.6} 
    & \rmse{1.03}{0.79}{1.27}
    \\ 
Jnk1\cite{szczepankiewicz_aminopyridinebased_2006,wang_accurate_2015}       & \notrust{2GMX }{-}\tnote{\jnkfig} & 0.77 
    & \begin{tabular}[t]{p{9em}} 
    \mildlytrust{3ELJ}{0.31}\tnote{\present}
      \end{tabular}
    & \begin{tabular}[t]{p{9em}} 
        \mildlytrust{3V3V}{1.5}\tnote{\similar, \ligdens, \packing}
      \end{tabular} 
    & \ligcount{21}  
    & \ligrange{3.4} 
    & \rmse{0.98}{0.68}{1.26}
    \\
MCL1\cite{friberg_discovery_2013,wang_accurate_2015}       & \hightrust{4HW3}{0.41}          & 0.26 
    & \begin{tabular}[t]{p{9em}} 
        \hightrust{6O6F}{0.30}   \\
        \hightrust{4ZBF}{0.35}\tnote{\similar} \\
        \hightrust{3WIX}{0.37}\tnote{\present, \lesssimilar}\\
      \end{tabular}
    & \begin{tabular}[t]{p{9em}}
        \hightrust{4WMU}{0.41}\tnote{\similar} \\
        \hightrust{4ZBI}{0.45}\tnote{\similar} \\	
      \end{tabular} 
    & \ligcount{42}  
    & \ligrange{4.2} 
    & \rmse{1.01}{0.80}{1.21}
    \\
P38(MAPK14) \cite{goldstein_discovery_2011,wang_accurate_2015} & \hightrust{3FLY}{0.6}           & 0.12 
    & \begin{tabular}[t]{p{9em}} 
        \hightrust{6SFI}{0.30}  \\
        \hightrust{3FMK}{0.30 }\tnote{\present, \similar}
      \end{tabular}
    & \begin{tabular}[t]{p{9em}} 
        \hightrust{3FLN}{0.33} \\
        \hightrust{3FMH}{0.43}\tnote{\present, \similar} \\
      \end{tabular} 
    & \ligcount{34}  
    & \ligrange{3.8}
    & \rmse{0.99}{0.76}{1.22}
    \\
PDE2\cite{buijnsters_structurebased_2014,perez-benito_predicting_2018}       & \mildlytrust{6EZF}{0.3}           & 0.07 
    & \begin{tabular}[t]{p{9em}} 
        \hightrust{6C7E}{ 0.29} \\
        \hightrust{5TYY}{0.30}\tnote{\present} \\
      \end{tabular}
    & \begin{tabular}[t]{p{9em}} 
        \hightrust{6B97}{0.46}\tnote{\lesssimilar}
      \end{tabular} 
    & \ligcount{21}  
    & \ligrange{3.2} 
    & \rmse{1.05}{0.75}{1.32}
    \\
PFKFB3\cite{boutard_discovery_2019,schindler_largescale_2020}     & \hightrust{6HVI}{0.31}          & 0.11 
    & \begin{tabular}[t]{p{9em}} 
        \hightrust{6HVH}{0.36}\tnote{\present, \similar}
      \end{tabular}        
    & 
    & \ligcount{40}  
    & \ligrange{3.8}
    & \rmse{1.04}{0.82}{1.25}
    \\
PTP1B\cite{wilson_structurebased_2007,wang_accurate_2015}     & \mildlytrust{2QBS}{0.33}          & 0.15 
    & \begin{tabular}[t]{p{9em}} 
        \hightrust{2HB1}{0.32}\tnote{\present} \\
        \mildlytrust{2ZMM}{0.33}\tnote{\similar, \crystal}
      \end{tabular}
    & \begin{tabular}[t]{p{9em}} 
        \hightrust{2QBR}{0.65}\tnote{\similar, \present}
      \end{tabular}
    & \ligcount{23}  
    & \ligrange{5.2}
    & \rmse{0.95}{0.67}{1.21}
    \\
SHP2\cite{chen_allosteric_2016,schindler_largescale_2020}       & \mildlytrust{5EHR}{0.32}          & 0.1  
    & \begin{tabular}[t]{p{9em}} 
       \mildlytrust{5EHP}{0.33}\tnote{\similar, \alternate} \\
       \hightrust{6MDD}{0.35}\tnote{\present}
      \end{tabular}
    & \begin{tabular}[t]{p{9em}} 
        \hightrust{6MD7}{0.35}
      \end{tabular} 
    & \ligcount{26}  
    & \ligrange{4.3}
    & \rmse{1.06}{0.76}{1.34}
    \\
SYK\cite{currie_discovery_2014,schindler_largescale_2020}     & \mildlytrust{4PV0}{0.69}
    & 0.19 
    & \begin{tabular}[t]{p{9em}} 
        \hightrust{4PX6}{0.3}\tnote{\present}
      \end{tabular}
    & \begin{tabular}[t]{p{9em}} 
        \hightrust{4FYO}{0.40}\tnote{\lesssimilar}
      \end{tabular}	
    & \ligcount{44}  
    & \ligrange{5.9}
    & \rmse{1.01}{0.81}{1.21}
    \\
Thrombin\cite{baum_more_2009,wang_accurate_2015}   &         \hightrust{2ZFF}{0.3}           & 0.06
    & \begin{tabular}[t]{p{9em}} 
        \hightrust{5JZY}{0.27}\tnote{\similar}
      \end{tabular}
    & \begin{tabular}[t]{p{9em}}
        \hightrust{3QX5}{0.28}\tnote{\present, \similar}
      \end{tabular}
    & \ligcount{11}  
    & \ligrange{1.7}
    & \rmse{0.97}{0.63}{1.28}
    \\
TNKS2\cite{buchstaller_discovery_2019}      
    & \hightrust{4UI5}{0.29}          
    & 0.08 
    & \begin{tabular}[t]{p{9em}} 
        \hightrust{4PC9}{0.27}\tnote{\present} \\
        \hightrust{4BU9}{0.29}\tnote{\present, \similar}
      \end{tabular}
    & \begin{tabular}[t]{p{9em}} 
        \hightrust{4UVZ}{0.29}\tnote{\similar} \\
    \end{tabular} 
    & \ligcount{27}  
    & \ligrange{4.4}
    & \rmse{1.04}{0.78}{1.28}
    \\
TYK2\cite{liang_lead_2013,liang_lead_2013a,wang_accurate_2015}       
    & \hightrust{4GIH}{0.5}
    & 0.15 
    & \begin{tabular}[t]{p{9em}} 
        \hightrust{3LXP}{0.31}\tnote{\present}
      \end{tabular}
    & \begin{tabular}[t]{p{9em}} 
        \mildlytrust{5WAL}{0.48}\tnote{\lesssimilar, \alternate}
    \end{tabular} 
    & \ligcount{16}  
    & \ligrange{4.3}
    & \rmse{0.97}{0.60}{1.30}
    \\
\hline
\end{tabular}
\begin{tablenotes}
\item [\present] structure was already available 6 month prior to publication of first benchmark study.%
\item [\similar] ligand similarity > 1.4.%
\item [\lesssimilar] ligand considerably similar > 0.8.%
\item [\crystal] crystal contacts.%
\item [\packing] packing.%
\item [\jnkfig] Fig. \ref{fig:jnk1_pdb_report}.%
\item [\alternate] alternate conformations.%
\item [\ligdens] low ligand density below 0.95.
\end{tablenotes}
}
\end{threeparttable}
\label{tab:struct}
\end{table*}

Any protein structural assessment should be done using two filters; overall (global) and local. Traditionally, overall quality of the structure (global) had been assessed using X-ray or cryo-EM resolution as it is easily accessible.
However, this metric provides a theoretical limit and does not assess the quality of the model. Therefore, it is not a good metric for accuracy, completeness or quality and should only be used alongside other metrics. Iridium, by design, does not set a resolution limit but suggests a resolution threshold of $< 3.5\,$\AA{}~\cite{warrenEssentialConsiderationsUsing2012} because it is difficult to model side chain atoms precisely above that threshold. Stricter thresholds have been suggested (i.e. $<2.0\,$\AA{}  in a recent benchmark~\cite{schindler_largescale_2020}).

More meaningful metrics for X-ray structures are $R$, $R_{\mathrm{free}}$ and the coordinate error. Currently, equivalent metrics for cryo-EM structures either do not exist or are less well understood.  As a result the rest of the discussion will focus on criteria for structures determined using X-ray or neutron diffraction data.  It should be noted that cryo-EM maps can still be visualized with the model to get an idea of the agreement between the model and the data.
The $R$-factor is a measure for the difference between the predicted data (by the model) and the measured data. A smaller $R$-factor indicates an experimentally consistent model. A complication with $R$-factor is that it is a non-normalized metric. For a given dataset the model with the lowest $R$-factor is best fit to the data.  Unfortunately, for different datasets, even for the same protein, lowest $R$-factor may not be the highest quality model. 
The $R_{\mathrm{free}}$-factor is calculated the same way, but uses only a held out randomly selected subset of the measured data. Thus, it can be used to identify overfit models as these result in a larger difference between $R$-factor and $R_{\mathrm{free}}$ (typically more than 0.05).
Both $R$-factors are easily accessible for reported crystallographic data, e.g. in the protein data bank (PDB)~\cite{bermanProteinDataBank2000}. 
The coordinate error, while more difficult to find or calculate, provides the best way to assess the precision and quality of the model:
\begin{equation}
    \mathrm{coordinate\ error} = \frac{2.22 R_{\mathrm{free}}\sqrt{N_i^3}\sqrt{V_a}} {n_{\mathrm{obs}}^{5/6}},
    \label{eq:coordinate_error}
\end{equation}
where $N_i$ is the number of heavy atoms with occupancy of 1, $V_a$ is the volume of the asymmetric unit cell and $n_{\mathrm{obs}}$ is the number of non-$R_{\mathrm{free}}$ reflections used during refinement. A high-quality structure should have a coordinate error $<0.7$. Recent PDB entries usually include a coordinate error estimate which can be found by searching for ESU $R_{\mathrm{free}}$, Cruickshank or Blow Density Precision Index (DPI). The coordinate error (as shown in Equation 1) is $\sqrt{3}\cdot \mathrm{BlowDPI}$. 

\begin{figure*}
    \centering
    \includegraphics[width=\textwidth]{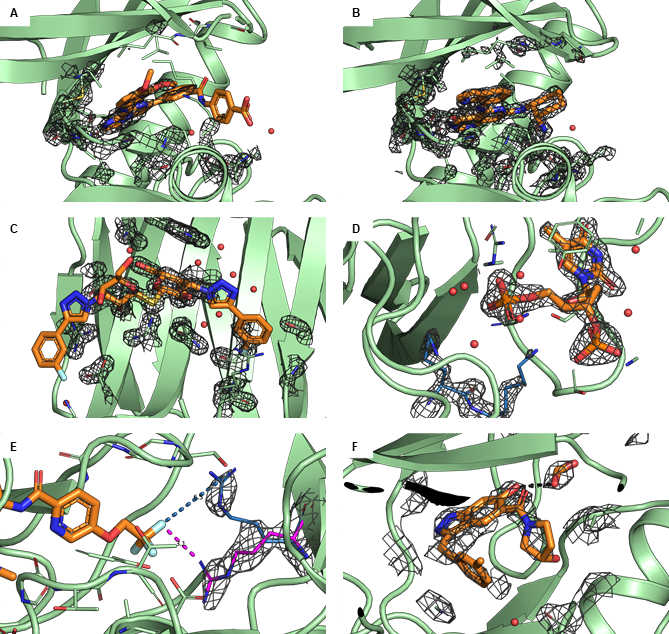}
    {
        \phantomsubcaption\label{fig:4pv0}
        \phantomsubcaption\label{fig:4px6}
        \phantomsubcaption\label{fig:5e89}
        \phantomsubcaption\label{fig:1snc}
        \phantomsubcaption\label{fig:3zov}
        \phantomsubcaption\label{fig:5hnb}
    }
    
    
    \caption{
    \textbf{Examples of common challenges encountered when using X-ray crystal structures.}
    The protein is shown in green and the ligand in orange. If not stated differently, the 2Fo-Fc maps are illustrated as grey isomesh at $2\sigma$ level. 
    \textbf{(\subref{fig:4pv0})} PDB ID 4PV0  shows poor density (at $3\sigma$) for residues in the active site.  The beta sheet loop at the top of the active site has residue side chains modeled with no density to support the conformation and the end of the loop has residues that are not modeled.
    \textbf{(\subref{fig:4px6})} The recommended structure PDB ID 4PX6 for the same protein  has complete density (and modeled atoms) for the whole loop (at $3\sigma$).
    \textbf{(\subref{fig:5e89})} PDB ID 5E89 shows poor ligand density, especially for the \textit{m}-Cl-phenyl (left) and the hydroxymethyl (center). This means that the ligand conformation, as shown, is not specified by the data, and thus should not be used as input to a computational study unless there is additional data supporting this binding mode.
    \textbf{(\subref{fig:1snc})} The ligand of PDB ID 1SNC has crystal contacts with the residues K70 and K71 (blue) of the neighboring unit that directly interact with the ligand, potentially affecting the binding mode relative to a solution environment.
    \textbf{(\subref{fig:3zov})} PDB ID 3ZOV has two
    alternate side chain conformations. Residue R368 in the B conformation (magenta) has clearly more density (0.75 $\sigma$) than the A conformation (blue). The B conformation interacts with the ligand (distance $3.2\,$\AA{}) whereas the A conformation does not interact with the ligand (distance $6.5\,$\AA{}). If the user does not look at both conformations and chooses A (by default), this would likely be incorrect and miss a potentially important protein-ligand interaction.  
    \textbf{(\subref{fig:5hnb})} In PDB ID 5HNB, there is an excipient (formic acid) that interacts directly with the ligand ($2.7\,$\AA{} O-O distance shown in black).  The formic acid could be replacing a bridging water.  From the data it is not possible to determine how the excipient is affecting the ligand/protein conformation, but for a study of ligand binding in the absence of formic acid, this should be removed.
    }
    \label{fig:crystal1}
\end{figure*}

While understanding the global quality of a structure is important, it is the local active site or ligand binding site that will have the largest impact on benchmarking performance. Therefore special care should be taken to assess the ligand and surrounding active site residues. 
Of highest priority is to identify all unmodeled residues and side chain atoms within 6~to~$8\,$\AA{} of any ligand atom. When multiple structures with similar coordinate error are available, the structure with no missing residues or side chain atoms that meets subsequent criteria should be used.
The electron density around the ligand should cover at least 90\% of the ligand atom centers, which can be checked visually or by checking for a real space correlation coefficient (RSCC) value $>0.90$. Examples for a poor ligand density are shown in Figures \ref{fig:4pv0} (in comparison to Figure \ref{fig:4px6} with a good density) and \ref{fig:5e89}. Ligand atoms where there are crystal packing atoms within $6\,$\AA{} should be identified (see Figure \ref{fig:1snc}), as such packing atoms may affect the observed binding mode. 
All ligand and active site atoms with occupancy $<1.0$ should be identified.
If there is only partial density for the ligand and the active site residue atoms, these partial-density atoms should be identified (see Figure \ref{fig:4pv0}). If alternate conformations for the ligand or active site residues are available, the selected conformation should be determined based on the electron density (see Figure \ref{fig:3zov}). Local metrics such as electron density support for individual atoms (EDIA)~\cite{meyder_estimating_2017} or a number of RSCC~\cite{tickle_statistical_2012} calculators can indicate if the electron density is sufficient to support the crystallographic placement of a given atom.
Covalently bound ligands should be identified and appropriately modelled.

\begin{figure}
    \centering
    \includegraphics[width=.47\textwidth]{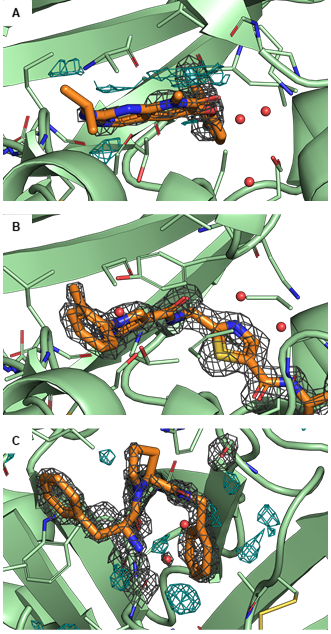}
    {
        \phantomsubcaption\label{fig:3fly}
        \phantomsubcaption\label{fig:6sfi}
        \phantomsubcaption\label{fig:2zff}
    }
    
    \caption{
    \textbf{Examples of challenges encountered for ligand modelling using X-ray crystal structures.}
    The protein is shown in green and the ligand in orange. If not stated differently, the 2Fo-Fc maps are illustrated as grey isomesh at $2\sigma$ level. In some panels, the difference density Fo-Fc map is illustrated as cyan isomesh at $+3\sigma$ level.
    \textbf{(\subref{fig:3fly})} In PDB ID 3FLY, there is significant difference density, likely indicating that the ligand conformation is not modeled correctly.  It is suspected that there is a low occupancy alternate conformation that is not modeled.
    \textbf{(\subref{fig:6sfi})} The suggested alternate structure of the same protein, PDB ID 6SFI, has no difference density.
    \textbf{(\subref{fig:2zff})} PDB ID 2ZFF shows unexplained electron density in the binding pocket (difference map, bottom, center, cyan). This could
    be either water or a Na$^+$ ion, as Na$^+$ is present and modeled in other sites. 
    }
    \label{fig:crystal2}
\end{figure}

Additional aspects should be considered beyond the quality of the model and the data (see also structure preparation, Section \ref{sec:prep}).
The structure of a complex could be deformed due to crystal contacts
or by experimental conditions like additives, pressure or temperature. These conditions might not be representative for the biological environment and therefore biologically active conformation of the complex (see Figure \ref{fig:1snc}). 
%
Other factors could play an important role in determining  active conformations, such as crystal waters, co-factors or co-binders. These should usually be included to model the natural environment of the protein (see Figure \ref{fig:5hnb} and \ref{fig:2zff}). One, however, needs to be cautious when retaining crystallographic waters in the binding site: for the cases where a modelled ligand clashes with the x-ray water, a careful equilibration with position restraints on the ligand atoms may be necessary to ensure further stable simulations. It may be undesired to trap waters near the bound ligand, as overhydration of the binding site may be detrimental to the free energy prediction accuracy due to potentially slow exchange of water between the binding cavity and the bulk~\cite{khalak2021absolutedg}. It is also important to remember that for X-ray data, modeling water (versus amino acids or organic compounds) is less precise than for other atoms particularly when the crystal is formed in a high salt environment.
The ligand in the experimental structure should be sufficiently close to the ligand to be simulated to have a model of the correct binding mode.

The criteria for selecting high-quality protein-ligand structures are summarized in the checklist "Choose Suitable Protein Structures for Benchmarking". A use case for these selection criteria to score and select structures from prior benchmarking datasets is found in Table \ref{tab:struct}.

A choice of the simulation conditions like temperature, ion concentration, other additives like co-factors or membranes require additional considerations. Ideally, these conditions are close to those for the structural experiment, the affinity measurements and physiological conditions. Most likely, a trade-off between all of these has to be found. Where possible select structures where data was collected at room temperature that were crystallized using non-salt precipitants. Be aware that room temperature data will have lower precision and more conformational heterogeneity.

If these requirements are not met, it does not necessarily mean that the data is not usable and the results will not match the experimental measurements. A structure not meeting the requirements may suffice after more manual intervention by the user, ideally an experienced one. Unresolved areas can be modelled with current tools and knowledge about atom interactions, though this can be a cause for concern if these are near the binding site. This concern has been validated, at least anecdotally, in a recent publication where different protein preparation procedures where shown to have a substantial effect on the accuracy of the free energy predictions~\cite{shih_impact_2020}.

Collective intelligence could be a way to mitigate the influence of individuals on the prepared input structures of a benchmark set. On a platform, other scientists could suggest changes to structures and updated versions could be deposited, increasing the quality of the benchmark set. Endorsement and rating of deposited structures could increase the level of trust given to specific structures and the database in general.

\subsection{Experimental binding affinity data}
\label{sec:affinities}
Choosing high-quality experimental data is crucial for constructing meaningful benchmarks of methods that predict ligand binding affinities. 
Evaluating whether experimental data merits inclusion requires an in-depth understanding of the biological system and the particular experimental assay that assesses protein-ligand affinity.
While a detailed overview of all experimental affinity measurement techniques is beyond the scope of this review, this section aims to summarize general aspects that should be considered when evaluating whether an experimental dataset is suitable for benchmarking purposes. 
We note that, in practice, it is often difficult to identify datasets that meet all the recommendations discussed below.

Overall, it is necessary that the experimental data used in benchmarks intended to measure the accuracy of reproducing experimental data are consistent, reliable, correspond well to the model system that is used in the simulations, allowing robust conclusions on accuracy to be drawn.

\subsubsection{Deriving free energies from experimental affinities}
Binding of a ligand to a receptor protein can be described as an equilibrium between unbound and bound states with the equilibrium constant of the dissociation $K_\mathrm{d}$ as
\begin{equation*}
    K_\mathrm{d} = \frac{[P][L]}{[PL]},
\end{equation*}
with $[PL]$ being the concentration of the bound protein-ligand complex and $[P]$ and $[L]$ the concentrations of the unbound protein and unbound ligand respectively. The binding free energy $\Delta G$ can be related to the dissociation constant via the following equation
\begin{equation}
    \label{eq:kdtodg}
    \Delta G = k_\mathrm{B} T \ln \frac{K_\mathrm{d}}{c^{\standardstate}},
\end{equation}
with Boltzmann constant $k_\mathrm{B}$, temperature $T$ and standard state reference concentration $c^{\standardstate}$, which is typically $c^{\standardstate}=1\,\mathrm{M}$. 
In many drug discovery projects, potency of compounds is assessed by measuring the half-maximal inhibitory concentration (IC$_{50}$) of a substance on a biological or biochemical function. This is often converted to pIC$_{50}$
\begin{equation*}
    \label{eq:pic50}
    \mathrm{pIC}_{50} = - \log_{10} \mathrm{IC}_{50}.
\end{equation*}
Typically, the substance is competing in these experiments with either a probe or substrate. For such competition assays, IC$_{50}$ can be related to the binding affinity of the inhibitor $K_i$ via the Cheng-Prusoff equation
\begin{equation}
\label{eq:CP}
    K_i = \frac{\mathrm{IC}_{50}}{1+\tfrac{[S]}{K_m}},
\end{equation}
where $[S]$ is the concentration of the substrate and $K_m$ the Michaelis constant. 
Assuming that all binding events result in effective protein inhibition, we can relate $K_i \approx K_d$.
Many assays are conducted using a substrate concentration of $[S] = K_m$. This leads to a conversion factor of $0.5$ between IC$_{50}$ and $K_i$ based on Equation~\ref{eq:CP} and to a constant offset in $\Delta G$. This offset cancels out for a congeneric ligand series with the same mode-of-action in identical assay conditions. Hence, in this case, $\Delta $pIC$_{50}$ values are a useful bioactivity that can be compared to relative binding free energy calculations. We can then use the approximation
\begin{equation*}
    \label{eq:ic50todg}
    \Delta\Delta G \approx k_B T \ln \frac{\mathrm{IC}_{50,b}}{\mathrm{IC}_{50,a}}.
\end{equation*}
For absolute $\Delta$G calculation comparison to experiment, the offset remains relevant. One suggestion to circumvent the issue could be to transform absolute $\Delta$G estimates to $\Delta\Delta$Gs, this way cancelling the offset and basing the further benchmarking on the relative free energy differences.

\subsubsection{Consistency of datasets}
\begin{figure}[!ht]
    \centering
    \includegraphics[width=0.95\linewidth]{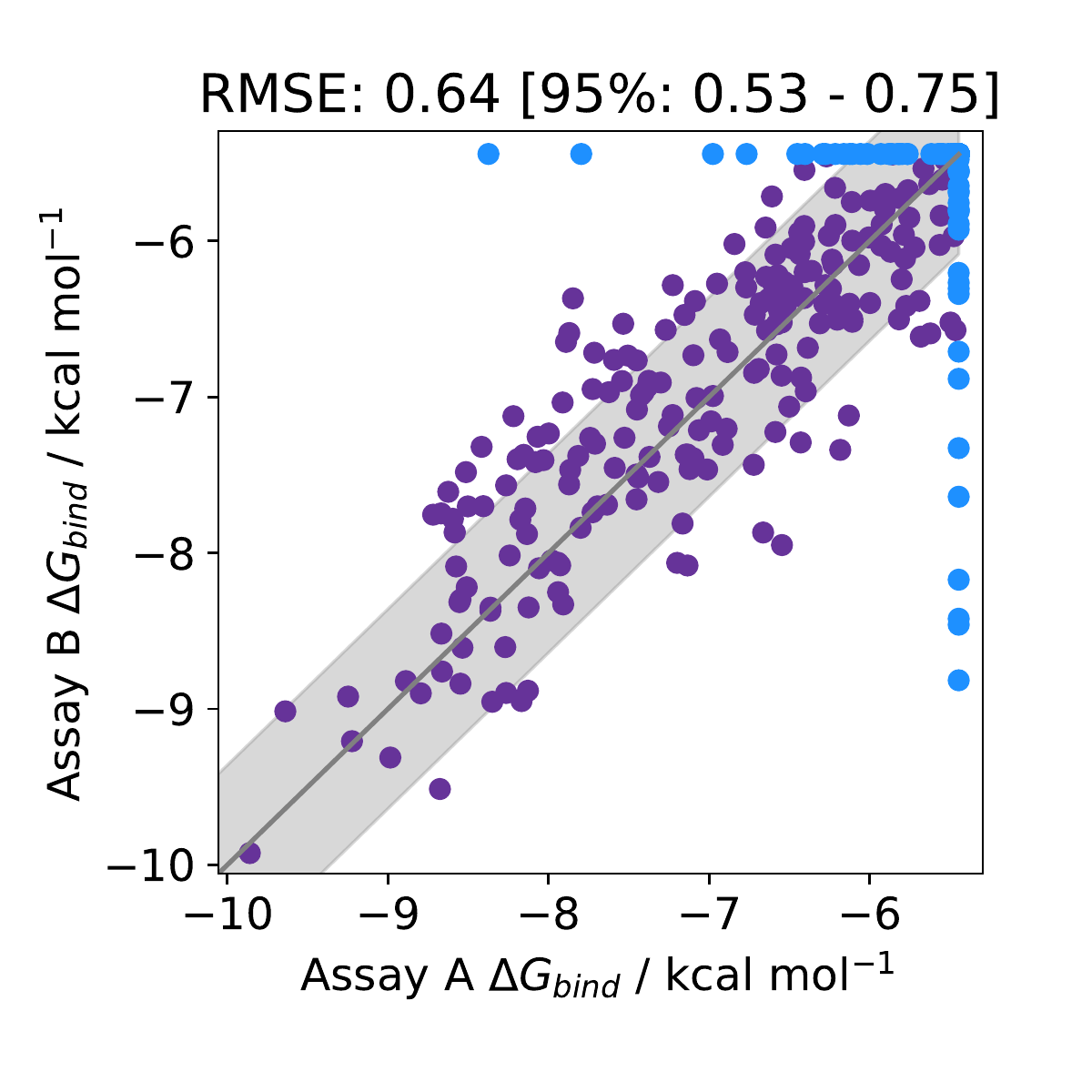}
    \caption{\textbf{Experimental uncertainties can be on the order of 0.64 kcal mol$^{-1}$.} The binding affinity of 365 molecules assayed by two different methods for the open source COVID moonshot project~\cite{achdout2020covid}. Molecules that were predicted to bind in one assay, but inactive (i.e., affinity lower than the assay limit) in the other are shown in blue. The RMSE agreement between the methods, for both purple and blue data points is 0.64 kcal mol$^{-1}$. Data was collected from the PostEra website~\cite{posteracovid} accessed 22/11/2020. The grey region indicates an assay variability of 0.64 kcal mol$^{-1}$.}
    \label{fig:expt_agreement}
\end{figure}

The paucity of experimental affinity measurement data may tempt practitioners to cobble together all available measurements for a given target (say, from a ChEMBL query) to construct a dataset with a sufficiently large number of measurements to provide statistical power in discriminating the performance of different methodologies on a given target.
This temptation should generally be resisted, as assay conditions or protocols in different labs might not be comparable. Figure \ref{fig:expt_agreement} illustrates this by comparing two sets of data obtained by different methods.
These differences could, for example, result from the concentration of the substrate (see Equation~\ref{eq:CP}), the protein construct, the incubation time or the composition of the buffer, and might not be sufficiently documented in the reported experimental methodology. However, in comparison to the inherent experimental error (see below), mixing experimental data from different laboratories might add only a moderate amount of noise~\cite{kalliokoski_comparability_2013}.
To ensure consistency within a dataset such that relative free energy differences are as reliable as possible, we highly recommend the use of data from a single source (e.g., a single publication or a patent).

To avoid rounding or unit conversion errors that often arise from automated or manual data extraction, data should be extracted from the original source.\footnote{Excellent examples of significant errors that can be introduced are thoroughly described in this comprehensive United States Geological Survey report on errors in misreporting the solubility and partition coefficient of dichlorodiphenyltrichloroethane (DDT) and its primary metabolite~\cite{pontolillo2001search}, as well as this talk on automatic data extraction errors~\cite{daga_pankaj_r_2019_3445476}.}
Going back to the original publication is also important to identify compounds that are outside of the detection limit of the assay but are still reported with specific numerical values (e.g., reported IC$_{50} > 30 \,\,\mu$M). Such ligands should be excluded from benchmark sets to ensure that accuracy measures can be properly evaluated.

\subsubsection{Experimental uncertainty}
\label{sec:exp_uncertainty}
To assess the reliability, ideally, errors are reported for all ligand affinities or at least for a subset. The primary publication of the experimental results is typically the best source of experimental uncertainty as cited affinities may occasionally be subject to rounding differences or unit errors~\cite{kramer2012experimental}. Errors quoted will likely be an estimate of the repeatability of the assay, rather than true, independent reproducibility. Publications with essential experimental controls reported --- such as incubation time and concentration regime to demonstrate equilibrium --- can add confidence to the reported affinity, however these may be performed and not reported~\cite{jarmoskaite2020measure}. Meta-analyses of both repeatability~\cite{sheridan2020experimental} and reproducibility~\cite{kramer2012experimental} found errors in pKi of 0.3-0.4 log units (0.43-0.58 kcal mol$^{-1}$) and 0.44 log units (0.64 kcal mol$^{-1}$) respectively.  Another analysis for reproducibility found that variability in pIC$_{50}$ was even 21-26\% higher than for pKi data (0.55 log units)~\cite{kalliokoski_comparability_2013}. These values provide a guideline for experimental error, if none is available. Note that for difference measures $\Delta $pIC$_{50}$, the individual experimental errors propagate as $\sqrt{\sigma_1^2+\sigma_2^2}$.

\subsubsection{Choosing representative experimental assays for FE calculations}

There are two main requirements to consider in order to ensure that the experimental data are representative of the physics-based binding free energy that is calculated from the simulations. First, the measured output should reflect or closely correlate with actual protein-ligand \emph{binding}. Second, the assay conditions and the protein-ligand system used in the simulation should match as closely as possible. The first point relates to choosing the appropriate type of experimental data to compare with. Ideally, these would be biophysical binding data such as $K_D$ determined from isothermal titration calorimetry (ITC) or surface plasmon resonance (SPR). However, this type of data is often only available for a small number of compounds in drug discovery projects (and the related literature), typically for a few representatives per series. In addition, ITC data are often only available for a narrow dynamic range~\cite{wiseman_rapid_1989,chodera_entropyenthalpy_2013}. Since having a sufficiently large dataset with a large dynamic range is also very important (see below), it may often be necessary to use data from functional assays (e.g., IC$_{50}$ from a biochemical assay) instead. For this assay, correlation with a biophysical readout should be checked before using the system as a benchmark dataset~\cite{kalliokoski_comparability_2013}.

With regards to matching simulation and binding assay, as mentioned above, it is important to have detailed knowledge of the assay conditions available; e.g., salt concentrations and co-factors. This information is needed for setting up a simulation model that closely matches the experimental conditions (see Section \ref{sec:prep}). Generally, salt concentration should match experimental assay conditions to capture screening effects, though sometimes salt \emph{identity} may be varied because of force field limitations. For a benchmark set, experimental data with assay conditions involving many co-factors or multiple protein partners should be avoided. In addition, one should check which protein construct was used in the structural studies compared to the assay (see Section \ref{sec:struct_data}). These should match as closely as possible. 

\subsubsection{Ensuring sufficient statistical power}
Finally, a dataset used for benchmarking of free energy calculations needs to be suitable to draw robust conclusions on the success of the methods ideally by both accuracy and correlation statistics. Whether a dataset is suitable depends on the number of data points in the set, the experimental dynamic range and the experimental uncertainty. 

\begin{figure}[!ht]
    \includegraphics[width=0.95\linewidth]{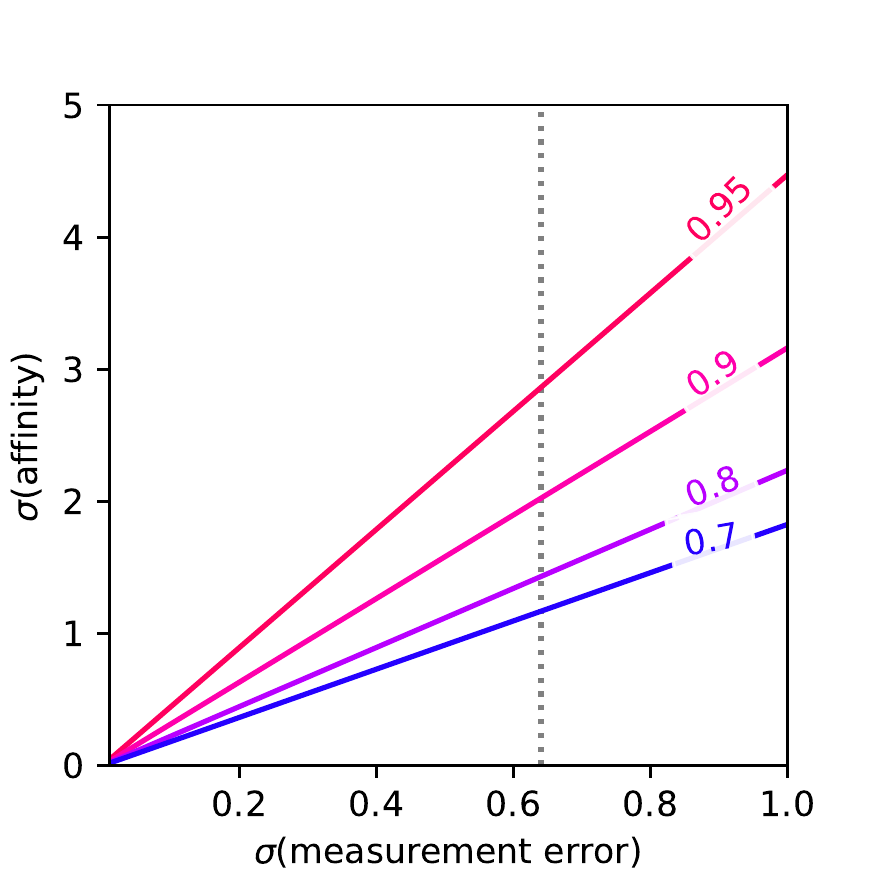}
    \caption{\textbf{The larger the experimental uncertainty, the larger the affinity range required for a given $R^2_{max}$}. Corresponding to Equation \ref{eqn:r2max}, the maximum achievable $R^2$ for a given dataset is limited by the range of affinities and the associated experimental uncertainty. The illustration assumes that $\sigma(\mathrm{measurement\ error})$ and $\sigma(\mathrm{affinity})$ are in the same units, with an experimental error of $0.64\,\mathrm{kcal\,mol^{-1}}$ indicated.}
    \label{fig:r2max}
\end{figure}

Quantifying the experimental uncertainty is necessary for understanding the upper-limit of feasible accuracy for a model~\cite{brown2009healthy}. Understanding this is both useful for fair comparison between methods, and for conveying the reliability of a model to medicinal chemists~\cite{griffen2020chemists}. Building predictive models becomes more difficult with (a) a small experimental dynamic range and (b) large experimental uncertainties. It is useful to understand the upper limit of success a computational method can have for a set of experimental results:
\begin{equation}\label{eqn:r2max}
    R^2_{\mathrm{max}} = 1 - \left(\frac{\sigma(\mathrm{measurement\   error})}{\sigma({\mathrm{affinity}})}\right) ^2,
\end{equation}
where $R^2_{\mathrm{max}}$ is the highest achievable $R^2$ for a dataset with a standard deviation of affinities ($\sigma(\mathrm{affinity})$) and an experimental uncertainty of  $\sigma\mathrm{(measurement\ error)}$~\cite{sheridan2020experimental}. This relation is illustrated in Figure \ref{fig:r2max}.

For a typical experimental error of 0.64 kcal mol$^{-1}$ (see Section~\ref{sec:exp_uncertainty}) and a desired $R^2_{\mathrm{max}} = 0.9$, a standard deviation of affinities $\sigma(\mathrm{affinity}) = 2.02 $ kcal mol$^{-1}$ ($\approx$1.5 log units) is required. Assuming a uniform distribution of experimental affinities in the dataset, this corresponds to a required dynamic range of 7.01 kcal mol$^{-1}$ (e.g., from $-12$ to $-5$ kcal mol$^{-1}$) or $\approx$ 5 log units (e.g., from 1 nM to 100 $\mu$M). This dynamic range and the associated standard deviation of affinities also allow to differentiate typical free energy methods from a trivial affinity prediction model where all predicted affinities $\Delta G_{\text{pred}}^i$ are equal to the mean experimental affinity $\sum_{i=1}^{N} \Delta G_{\text{exp}}^i$. Note that for such a model RMSE is equal to the standard deviation of the affinities $\sigma({\mathrm{affinity}})$, while there is no correlation between predicted and experimental affinities. In practice, experimental datasets with a dynamic range of $7\,\mathrm{kcal\,mol^{-1}}$ are difficult to obtain. Using the same assumptions as before, a dynamic range of 5 and $3\,\mathrm{kcal\,mol^{-1}}$ correspond to a standard deviation of affinities of $\sigma(\mathrm{affinity}) = 1.44\,\mathrm{kcal\,mol^{-1}}$ and $\sigma(\mathrm{affinity}) = 0.87\,\mathrm{kcal\,mol^{-1}}$ and hence $R^2_{\mathrm{max}} = 0.8$ and $R^2_{\mathrm{max}} = 0.45$ respectively. Balancing data availability and achievable $R^2_{\mathrm{max}}$, we recommend collecting datasets with a dynamic range of $5\,\mathrm{kcal\,mol^{-1}}$.

\begin{figure}[!ht]
    \centering
    \includegraphics[width=0.95\linewidth]{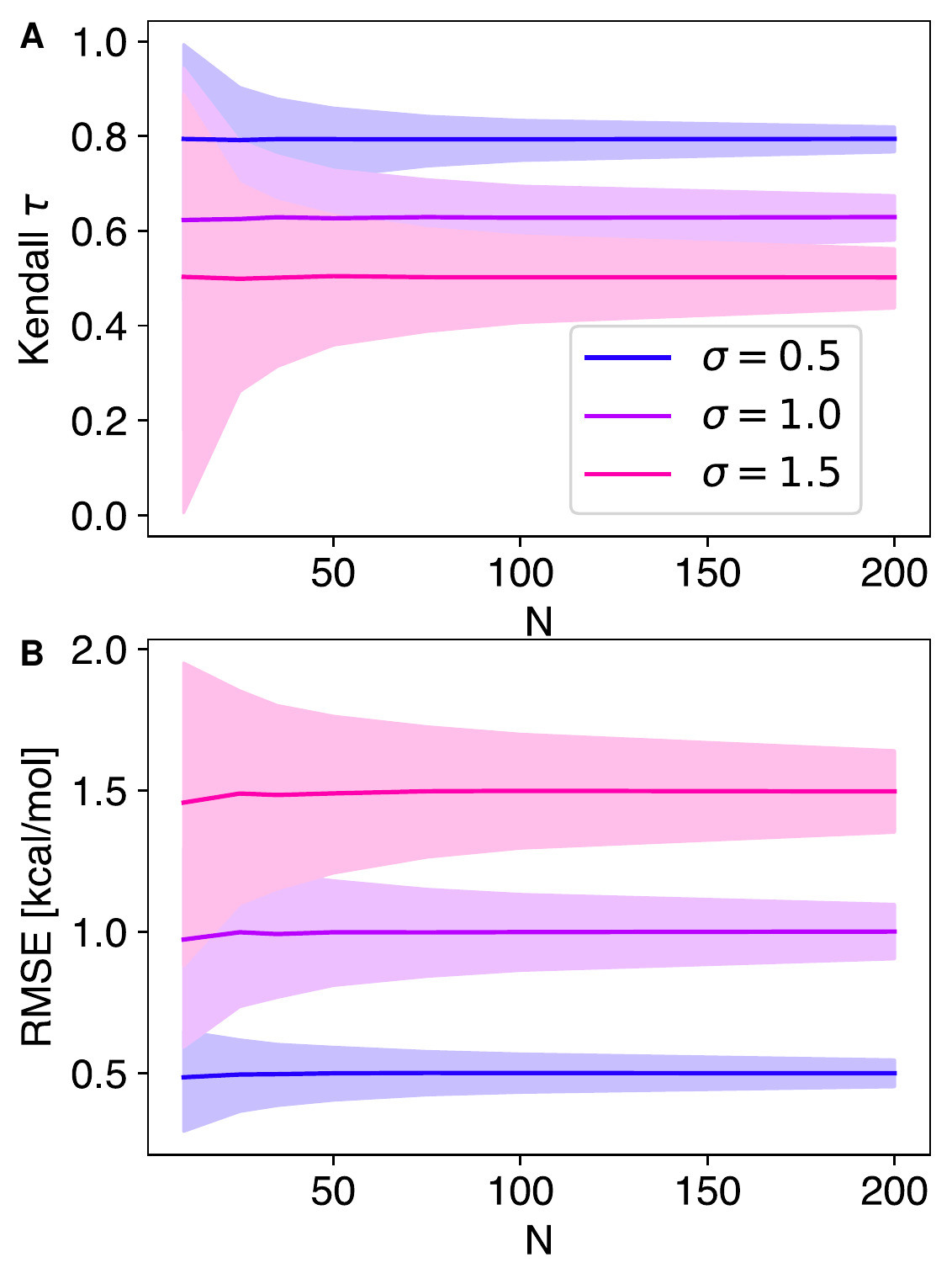}
    \caption{\textbf{The larger the dataset, the smaller the uncertainty in the performance statistics}. (\textbf{A}) Kendall $\tau$ and (\textbf{B}) RMSE were evaluated for 1,000 toy datasets for a given size of the dataset $N$. The experimental data were simulated from a uniform distribution over the interval [-12:-5] and the predicted affinities were simulated from the experimental toy data using a Gaussian distribution with different standard deviation $\sigma$. The statistic was evaluated for the whole dataset and 95\% confidence intervals were estimated via bootstrapping. These were then averaged over all 1,000 toy datasets.}
    \label{fig:N_CI}
\end{figure}

In order to robustly evaluate statistics with small confidence intervals, the dataset needs to be sufficiently large. 
Figure \ref{fig:N_CI} illustrates the dependence of the confidence interval obtained by bootstrapping for correlation statistics and accuracy statistics for simulated toy data. 
The "experimental" toy data were simulated using a uniform distribution with an affinity range of $7\,\mathrm{kcal\,mol^{-1}}$. 
This would be the optimal dynamic range for an experimental error of $0.64\,\mathrm{kcal\,mol^{-1}}$ (see Section~\ref{sec:exp_uncertainty}). 
Predicted toy data were derived from the experimental toy data using a Gaussian distribution with standard deviation of $\sigma = 0.5$, 1 and $1.5\,\mathrm{kcal\,mol^{-1}}$. 
While the absolute values that can be obtained for the correlation statistics are strongly affected by the dynamic range of the experimental data, the effect on the confidence intervals estimated via bootstrapping is relatively small (very similar results in terms of the size of the confidence intervals can be obtained assuming a dynamic range of $5\,\mathrm{kcal\,mol^{-1}}$).

Based on these simulations, we recommend a dataset size of 25 to 50 ligands. For a dataset size of 50, it is possible to distinguish between all three toy methods reliably in terms of RMSE. For an affinity prediction method with Gaussian error $\sigma = 1.0\,\mathrm{kcal\,mol^{-1}}$ this would yield the following estimated statistics: Kendall $\tau = 0.72_{0.62}^{0.80}$ and RMSE $= 1.0_{0.81}^{1.18}\,\mathrm{kcal\,mol^{-1}}$. Note that for relative calculations, a smaller number of ligands could be sufficient since multiple edges are typically evaluated for each ligand. On the other hand, for relative calculations, the experimental error for the relative free energies are larger because experimental errors for both ligands add up. 

As demonstrated in the example above and Figure~\ref{fig:N_CI}, uncertainties of the estimated statistics strongly depend on the standard errors of individual free energy predictions. Naturally, this necessitates accurate estimation of uncertainties for the calculated $\Delta$G (or $\Delta\Delta$G) values. Depending on the free energy protocol and $\Delta$G estimator used, an analytical uncertainty estimator might be available. Another possibility is to use bootstrapping, i.e. resample the raw calculation data with repetition to reconstruct the sampling distribution and estimate its standard deviation. However, probably the most reliable, yet computationally demanding, approach to obtaining the standard error is by repeating the whole calculation procedure multiple times.~\cite{knapp2018replicas,gapsys2020replicas,wan2021uncertainty}

As stated before, in practice it is challenging to find datasets that meet these criteria for dynamic range and number of ligands. We therefore currently recommend annotating benchmark datasets according to these criteria to make challenges and limitations visible.

\section{How to best set up and run benchmark free energy simulations}

\subsection{Structure preparation}
\label{sec:prep}

Starting with an experimental crystal structure, often an X-ray structure for the protein or protein-ligand complex, the most error-prone stage of protein preparation is the translation from this experimental structure into a simulation model: inferring missing atoms and making choices about which X-ray components to include. Having chosen biological unit based on the criteria in the above section, some domains of the structure may be removed if they are large and unlikely to affect the biological activities of interest. The truncation of the system needs to be assessed carefully as it has been shown in some cases, such as the dimeric form of PDE2 and the presence of cyclin with CDK2, as a more authentic representation of the system was beneficial for stability during simulations and improved the free energy calculations. In some cases, though, truncation gains efficiency by decreasing the size of the overall simulation system while maintaining its biological activity, with potentially minimal impact on results. Datasets for benchmarking may be run many times so this efficiency gain can be meaningful.

In addition to the protein itself, the subsystem carried forward from the X-ray structure into simulation may have other components: ligand, cofactors, structural waters, other ligands (if simulating a multimer), post-translational modifications (PTMs), and excipients. The cofactors should be deliberately included or excluded based on their role in the biological activity being modeled, removing a cofactor from its cavity might cause unexpected movements or collapse of the cavity during the simulations. To avoid this, a careful equilibration and solvation of that pocket might be needed. All structural waters close to the protein should be considered for inclusion by carefully verifying that water positions are compatible with the modelled ligands: in principle the MD sampling could allow waters to arrange in equilibrium positions, but experimental and theoretical work has shown that the timescales for this can be impractically long. Also, internal structural waters even very distal from the active site are integral to the protein structure, and omitting them can adversely affect the protein dynamics. Generally, we recommend excluding excipients (often specific to the crystallization media and not present in the assay). PTMs require a judgement call: surface-exposed and distal from the active site they can often be safely excluded, for example glycosylations which could otherwise greatly increase the size of the calculation. This again can save on the overall system size and prevent parameterization difficulties. PTMs proximal to the active site or known to be directly implicated in activity should be retained. Ligands other than that in the active site are again a judgement call: retaining them is only necessary if there is biological cooperativity in the biological assay. As this is in practice often not known, they should be kept if possible. 

For the absolute binding free energy calculations, it is also necessary to account for the free energy change in the protein's transition from its apo to holo state. Therefore, initializing the simulations of the alchemical ligand coupling based on the crystallographically resolved protein's apo state may facilitate convergence.

\begin{figure}[!ht]
    \centering
    \includegraphics[width=0.95\linewidth]{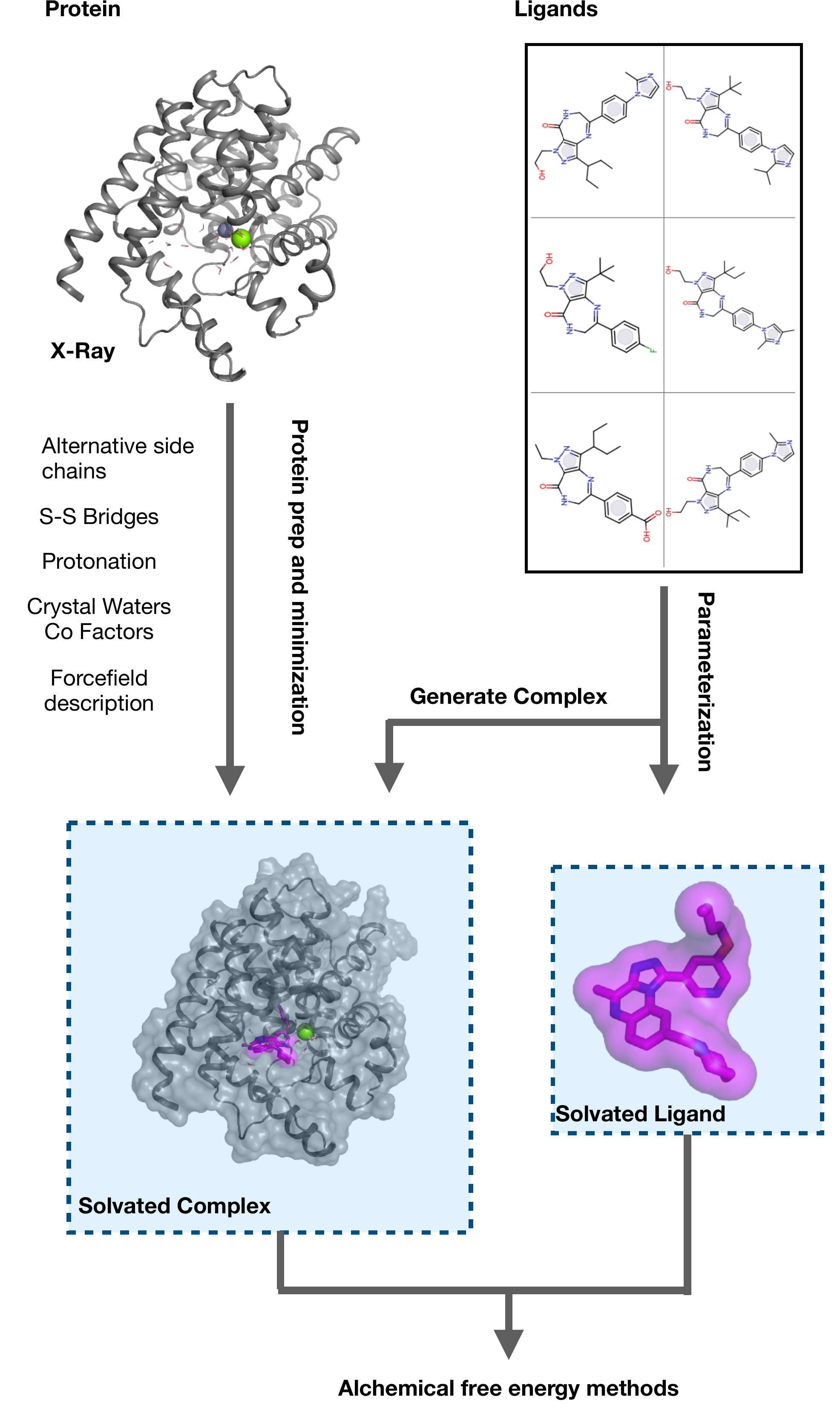}
    \caption{\textbf{Outline of the system preparation steps.} 
    First the protein is prepared (left, Section \ref{sec:protprep}) by modelling missing atoms, assigning bond orders, protonation and tautomeric states. 
    Similarily, the chemical structure of the ligands is translated into a simulation model (right, Section \ref{sec:ligprep}). 
    The ligands are simulated in two different environments, once complexed with the protein (bottom left) and once in solvent (bottom right). 
    For the solvated complex, the ligand structures need to be docked into the binding site of the protein, typically by using the information of a reference ligand in the X-ray structure.
    }
    \label{fig:system_preparation}
\end{figure}

\subsubsection{Protein preparation}
\label{sec:protprep}

The experimental protein structure frequently has missing coordinates for atoms, residues or groups of residues due to the lack of supporting data (electron density) from the X-ray experiments. These often include N-terminal and C-terminal residues, mobile loops (e.g. the activation loop in kinases), and residue sidechains. Also, there can be extra coordinates available in the structure as "alternate locations" (AltLocs): residue sidechains, or occasionally entire residues or the ligand, for which the experimental density supports more than one distinct orientation in a single X-ray structure solution. For the simulation, the protein must have all the atoms provided for every residue modeled. Missing residue sidechains should always be modeled in, assigning them the most preferred rotamer given the local environment. 

If the N- and/or C-terminal residues are missing due to lack of electron density, this may provide a basis for omitting them from the model, but the truncated N- and C-termini should be "capped" by neutral termini, usually an acetate (ACE) cap on the N-terminus and an N-methyl (NME) cap on the C-terminus to mimic the peptide backbone up to the carbon-alpha. Of course, one must be careful not to cap the charged protein termini which are properly resolved in the X-ray: these can be critical for function and structure. 

This "capping" tactic can also treat the termini of "gaps": regions of missing residues over the span of the peptide chain, usually missing loop regions due to lack of experimental density. While capping the ends of a loop instead of modeling the whole loop may be acceptable for MD runs of relatively short duration, over longer simulations there is a risk of having the protein around the capped ends of the missing loop gradually lose its structure. Even if a loop is unstructured (and therefore not resolved in the X-ray structure), its presence still affects the remainder of the structure and can provide stability by restricting movement of the connecting residues, thus raising concerns if these are capped instead. Strategic use of a distance restraints during the simulations can mitigate this liability.

Another possibility for missing loops is to close the ends with a short modeled loop of glycine residues of sufficient size to link the termini without introducing strain, but not necessarily of the full length of the missing loop. There are several reasons why this can be desirable. If the missing loop is particularly large (for instance >15 or 20 amino acids) accurately modeling its conformation could be challenging and introduce more uncertainty and instability to MD simulations. Furthermore, if the missing loop is distal from the binding site and not expected to affect protein-ligand interactions, the replacement only needs to stabilize the termini and avoids the use of restraints. 

However, all of these approaches are likely inferior to using a good quality model of the missing loop. 

When multiple alternate models of a particular region of the protein are available, the experimental model indicates that this region potentially occupies two (or more) mutually exclusive conformations, but one must be chosen for the model. Again, this selection can be a judgement call depending on where this region occurs relative to the active site: distal from the active site, the choice may be less critical; proximal requires more careful consideration. Higher occupancy for one of the alternate models could provide a reason to choose that particular model for the calculations. For critical or uncertain cases, we recommend repeating test calculations beginning from different models to analyze the sensitivity of the choice. 

Once the above issues have been resolved, there remains one more round of decision-making to select sidechain rotamers and protonation states. Protein X-ray experiments usually cannot resolve the positions of hydrogens, making decisions on the protonation states an issue. Another challenge is determining sidechain orientations: sidechain flips are particularly relevant for HIS, ASN, and GLN, because the X-ray crystallography experiments cannot distinguish between different first-row elements O, N, and C. These elements produce similar density and are indistinguishable. This means that even with good electron density the sidechain orientations of ASN and GLN can have either orientation, swapping O and N positions, and thus interchanging H-bond donors and acceptors. The two possible orientations of HIS sidechains effectively interchange N and C positions in the ring. Surface exposed, these different orientations may be of little consequence, but in the interior of the protein, proximal to the active site, or especially interacting with the ligand, this can be very important and can change patterns of hydrogen bond donors and acceptors. In principle these orientations can be sampled over the course of the MD run but only if the trajectory is long enough for the sampling scheme to allow it. Considering that these orientations are experimentally ambiguous, it is a matter of judgement at setup time of whether these sidechains should be reoriented to make a more chemically reasonable model. 

Protonation of the protein model is generally straightforward with one key exception: the ionization state of sidechains, especially HIS, ASP and GLU, which may undergo pKa shifts due to changes in the environment. Active site catalytic CYS is another case requiring care, and occasionally LYS can be deprotonated in some circumstances. The two main determining factors are the pH of the biological milieu and the microscopic environment around the ionizable sidechain. In general, the ionization state of each residue is chosen during the setup of the protein and remains constant over the course of the simulation, even if the microenvironment changes. 
Note that a formal charge on the bound ligand can also affect the ionization state of nearby protein residues; this can be particularly problematic when the ligand charge alchemically changes over the course of a relative free energy calculation. Unlike side-chain rotamers, which may sample other orientations within a simulation, incorrect protonation state assignments cannot correct themselves without the use of constant-pH algorithms, that have not been routinely implemented within free energy calculations yet. 

There are a number of tools to automate the steps described in this section, notably the Protein Preparation Wizard~\cite{madhavisastry_protein_2013}, the Molecular Operating Environment (MOE)~\cite{moe_2021}, and Spruce~\cite{openeye_spruce}. We recommend manual inspections after applying these.

\subsubsection{Ligand preparation}
\label{sec:ligprep}

In the preparation of the ligand for simulation it is important to verify that the chemical structure is correct. While this is less problematic for structures generated from small-molecule sources, historically it has been a frequent problem for ligands taken from protein-ligand X-ray structures. Since X-ray structures lack protons and do not provide bond orders or other key information, if a PDB structure is used as input, some tool must be applied to supply this information, presenting a frequent source of failure (though, for structures in the RCSB, a ligand SMILES string can provide a more complete representation of the ligand's identity).

Once the underlying chemical structure, including bond orders and stereochemistry, is correct, the key issues are the  tautomer and ionization states. As with the ionizable protein residue discussed above, the main factors are the macroscopic pKa of the ligand (for ionization states), the intrinsic relative stability of different tautomer states, and the perturbing effects of the active site micro-environment of the bound ligand. Compounding the complexity is if the unbound ligand (used as a reference state) would have a different tautomer/ionization state. These need to be carefully examined at setup to make sure there is complementarity between the protein and ligand independently of the alchemical change between ligands, and then to flag and resolve alchemical conversions between inconsistent states of the protein.

\subsubsection{Preparation of the complex}

Once protein and ligand have been prepared, the complex is assembled and solvated in water with counter-ions at an appropriate ionic strength, or embedded in membrane if the protein belongs to a membrane protein family. Membrane simulations should use an appropriate equilibrated membrane that matches experimental criteria of thickness and area per lipid as well as the appropriate counter ions. Once the system box is constructed the step involves neutralizing the net charge on the protein-ligand complex, but beyond this a higher concentration of salt (usually sodium chloride) is often warranted to mimic the biological milieu being modeled; most assays are run in a significant salt concentration (100 to 150 mM) to emulate biological environments. The salt concentration can strongly affect experimental binding affinities, particularly with highly polar active sites. The ion placement needs to be handled with care, for example by prohibiting insertion of ions within a given distance from the protein-ligand complex. Otherwise, positioning an ion in a close proximity to the bound ligand may destabilize the binding pose, in turn affecting the prediction accuracy~\cite{aldeghi2018resistance}. 

Once the above decisions have been made and the complete simulation system has been set up, it is important to let it relax and equilibrate at simulation temperature and pressure, which should mimic the assay conditions.

\subsection{Alchemical free energy calculations pose specific setup challenges}
\label{sec:alchemical_prep}

There are an abundance of details that must be considered during the set up of any simulation and in particular for alchemical free energy calculations. These simulations require setting up an alchemical perturbation of the small molecules, but also require making a variety of  assumptions with respect to the environment at the two endstates. In the following we will address all essential choices that need to be made for the setup. For a very detailed introduction to best practices for alchemical free energy calculations and a much broader discussion on choices for their setup please refer to the relevant best practices guide~\cite{meyBestPracticesAlchemical2020}. 

\subsubsection{Should I run an absolute or relative free energy calculation?}
There are two possible ways in which to run alchemical free energy calculations, which both provide free energies of binding, but will require different routes for their setup. \textit{Relative} free energy calculations provide free energies of binding with respect to a reference ligand, meaning that all compounds that are to be assessed for their binding affinity should share a similar scaffold. In contrast, \textit{absolute} free energies of binding can be used for a set of ligands that do not share any commonalities, since the reference state for the free energy of binding is the standard state. This is probably the easiest deciding factor in terms of what kind of calculation to run. If the particular benchmark dataset contains ligands that form a congeneric series then a relative calculation is likely a better choice. Of course, congeneric ligand series can also be assessed using absolute free energy calculations, or it may be of interest to compare relative to absolute calculations for a given benchmark dataset.

\subsubsection{Alchemical pathway}

\paragraph{Choices in topology}
The choice of topology may be dictated by the simulation software of choice as not all common MD codes implement all topologies. The topology refers to the way in which a molecule A is changed to molecule B, in case of relative free energy calculations. Selecting either a dual or single topology approach is acceptable, unless performance of different topologies is assessed across the benchmark datasets. For more details on the different topology choices and implementations please refer to Mey et al.~\cite{meyBestPracticesAlchemical2020}.

\paragraph{Choices concerning $\lambda$}
In order to connect the initial and final state of the alchemical free energy calculation an alchemical pathway must be chosen. This pathway is regulated by a variable $\lambda$, which, in the simplest formulation, at $\lambda=0$ represents molecule A and at $\lambda=1$ molecule B. As free energy is a state function, the computed free energy is in principle independent on the pathway, but different choices in pathway can make the problem computationally more or less tractable. The simplest way to switch between molecule A and B is using a linear switching function for the Hamiltonian of the form:
\begin{equation}
\label{eq:switching}
\mathcal{H}(\vec{q},\vec{p},\vec{\lambda}) = (1-\vec{\lambda}) \mathcal{H}_0(\vec{q},\vec{p}) + \vec{\lambda}\mathcal{H}_1(\vec{q},\vec{p}),
\end{equation}
where $\mathcal{H}$ is the Hamiltonian, $\vec{q}$ is the set of positions, $\vec{p}$ is the set of momenta and $\vec{\lambda}$ the switching parameter. However, this typical approach fails when atoms are being inserted or deleted, requiring alternate choices, as reviewed elsewhere~\cite{meyBestPracticesAlchemical2020}.

For the free energy perturbation (FEP) protocol, considerable care needs to be taken in selecting the switching function and spacing of so-called $\lambda$-windows. Common choices are, how many $\lambda$-windows should be used? What functional form should my switching function take? The concept of \textit{difficult} and \textit{easy} transformation is more and more explored, but currently heuristics based on phase space overlap between neighboring $\lambda$-windows is the best way to assess how many windows should be simulated. This can for example be done by looking at the off-diagonals of an overlap matrix~\cite{klimovich_guidelines_2015,kuhn_assessment_2020}. Furthermore, the choice of simulation protocol will influence what switching function and how many $\lambda$-windows should be used. 

\subsubsection{Choice of simulation protocol}
\begin{figure}
    \includegraphics[width=0.95\linewidth]{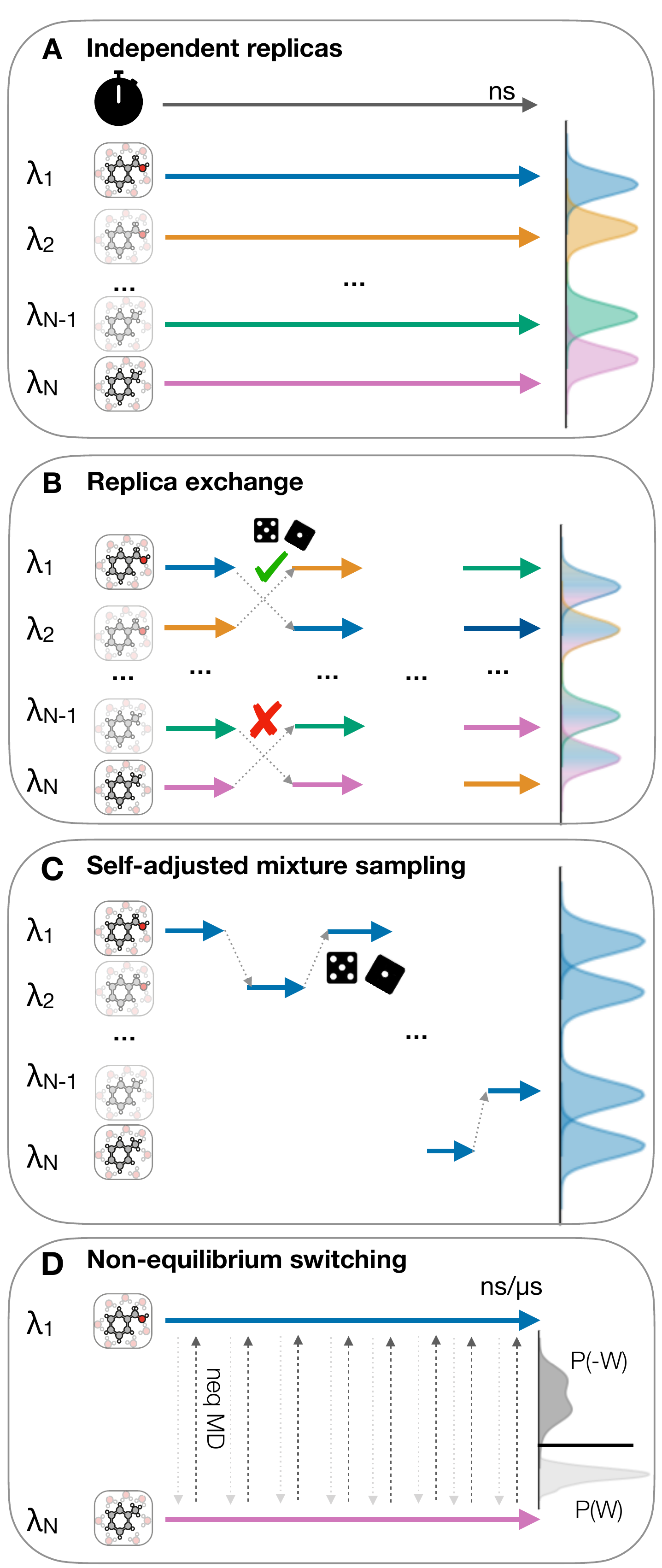}
    \caption{\textbf{There are four simulations protocols available for for generating samples and evaluating the Hamiltonian at the $\vec{\lambda}$ states.} \textbf{(A)} Independent replicas run in parallel at different $\lambda$s as indicated by differently colored arrows, \textbf{(B)} Replica exchange attempts after short simulation for each replica \textbf{(C)} Self-adjusted mixture sampling with a single replica exploring all of $\lambda$, \textbf{(D)} Non-equilibrium methods with equilibrium end-state simulation and frequent non-equilibrium switching between endstates. The clock icon is indicating the flow of simulation time and the pair of dice indicate a Metropolis Hastings based trial move}
    \label{fig:protocols}
\end{figure}

Various simulations protocols for alchemical free energy calculations are available and can be categorized into reference state, independent replica with constant or variable $\lambda$ states and ensemble (multiple replica) methods.
In reference state methods, one reference state is simulated in a single simulation and free energy differences to other states are extrapolated. Examples for these is one step perturbation~\cite{zwanzigHighTemperatureEquation1954,liu_estimating_1996,raman_sitespecific_2012,raman_estimation_2017,boresch_convergence_2017} or enveloping distribution sampling (EDS)~\cite{christ_enveloping_2007, christ_multiple_2008,christ_comparison_2009,sidler_efficient_2017,perthold_accelerated_2018}.
In independent replica methods, one or several simulations are performed at different states of the coupling parameter $\lambda$. These $\lambda$ parameters may be constant like in discrete thermodynamic integration\cite{kirkwoodQuantumStatisticsAlmost1933,kirkwoodQuantumStatisticsAlmost1934,kirkwoodStatisticalMechanicsFluid1935} or free energy perturbation\cite{jorgensen_perspective_2008}. Other methods allow the simulation to adopt discrete $\lambda$ states as in self-adjusted mixture sampling~\cite{tan_optimally_2017} or expanded ensemble simulations~\cite{lyubartsev_new_1992,lyubartsev_free_1994,lyubartsev_determination_1996,escobedo_optimized_2007,escobedo_optimized_2007a}. $\lambda$ can also be varied continuously in slow growth thermodynamic integration~\cite{straatsma_free_1986} or $\lambda$ dynamics~\cite{kong_dynamics_1996,guo_application_2003,knight_ldynamics_2009,knight_multisite_2011,donnini_constant_2011,armacost_biasing_2015,hayes_adaptive_2017}. Fast growth or non-equilibrium switching are special cases of independent replica methods where $\lambda$ is rapidly changed in non-equilibrium simulations ~\cite{jarzynski_nonequilibrium_1997,crooks_pathensemble_2000,hendrix_fast_2001,hummer_fastgrowth_2002}.
In multiple replica or ensemble methods, two or more replicas of the same system are simulated in parallel and are in equilibrium with each other. In Hamiltonian replica exchange, swaps between replicas at different fixed $\lambda$ states are attempted and either accepted or rejected according to the Metropolis-Hastings criterion~\cite{sugita_replicaexchange_1999,fukunishi_hamiltonian_2002,zhang_simulating_2016}.
We provide examples of four of the above protocols, which are summarised in Figure~\ref{fig:protocols}. These are: Figure~\ref{fig:protocols} (A) independent replicas at constant $\lambda$ states, (B) replica exchange, (C) Single replica, self adjusted mixture modelling and (D) non-equilibrium switching. Particularly for (B) and (C) the choice of $\lambda$-spacing will be important, as in (B) it dictates the success of replicas exchanging between $\lambda$s and in (C), often tightly spaced replicas allow for a best exploration. Independent replicas are not necessarily recommended, but are still commonly implemented in software packages. 

\subsubsection{End-state environments}
When setting up a relative free energy calculation it is important to be aware of the similarity of the 'end states', i.e. of the conformational, hydration, and electrostatic environment of ligand A and B. Many of these end-state issues can be addressed with longer sampling, but this may be impractical and should be considered when planning perturbations. Issues can arise, if there are two distinct bound conformations (different binding modes) for ligand A and ligand B, it may be necessary to sample both binding modes, or extend the simulation time to allow for sufficient rearrangements. A similar issue that may be addressed with extended sampling times are scaffold changes that occur between ligand A and B. Different hydration patterns may also cause inaccuracies in computed binding free energies. Probably the most difficult issue to address are changes in charge states that occur either between the two ligands or may even affect the protein depending on the type of ligand binding. Methods to address this issue are double system/single box setups~\cite{gapsys2015dssb} to retain neutral charges, the use of alchemical ions~\cite{chen2018chargecorrections}, or the post-hoc corrections~\cite{rocklin2013chargecorrection,reif2014chargecorrections} to $\Delta G$ values.

For the absolute binding free energy calculations, the situation is further complicated by the need to account for the $\Delta$G change in protein's conformation when transitioning from apo to holo state. Converging larger protein reorganizations may become challenging already in relative free energy calculations~\cite{lim2016sensitivity}, thus in estimating absolute binding $\Delta$Gs failure to properly capture this contribution may manifest in a substantial offset of predicted values with respect to experimental measurement~\cite{khalak2021absolutedg}. In principle, longer or enhanced sampling could help in improving convergence of large conformational changes. Another option is to explicitly make use of the crystallographic apo state (if it is available) to initialize ligand coupling simulations for the non-equilibrium switching scheme~\cite{gapsys2021absdg} or seed an FEP based simulation~\cite{hahn2020FEPseeding}.

\subsubsection{Perturbation maps for relative calculations}
In relative free energy calculations a network of perturbations between ligands needs to be constructed. The choice of which relative calculations to carry out is vast and can have a substantial effect on the accuracy of the results. The way in which different ligands are connected by relative alchemical calculations is called a perturbation map. In particular for benchmarking free energy methods, perturbation maps should be held fixed for a given benchmark set, unless the goal is to test different approaches for setting up perturbation maps. In this way each edge of the perturbation map will be maintained across subsequent tests and plots created during the analysis phase later will be comparable. 

\begin{figure}[!ht]
    \centering
    \includegraphics[trim=0 0 200 0, clip, width=0.9\linewidth]{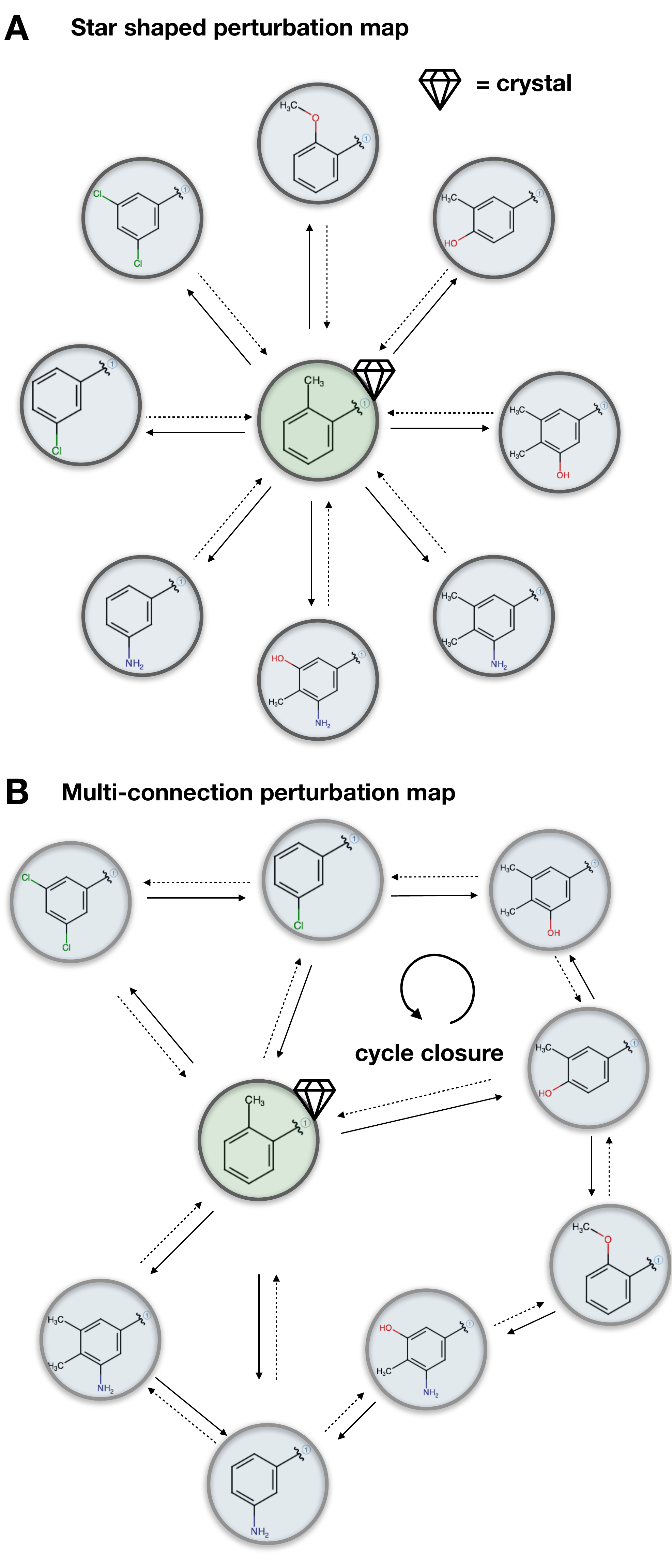}
    \caption{\textbf{Typically either star shaped perturbation maps or multi-connected perturbation maps are used in relative free energy calculations.} \textbf{(A)} The star map will have a central ligand, of which the crystal structure is known and all other ligands distributed in a star. \textbf{(B)} A multi-connected map introduces redundancies into the network, allows for larger perturbations through multiple connections and allows assessment of robustness of calculations. The diamond and green shading indicates the crystal structure.}
    \label{fig:map}
\end{figure}

The simplest way of connecting ligands in a perturbation map is in a star shape, with each connected to a central crystal structure ligand, with the assumption that all ligands of the congeneric series will bind in the same binding mode as the available crystal -- which may even be confirmed by other crystals, see Figure~\ref{fig:map} (A), there are different methods available for creating interconnected perturbation maps using LOMAP~\cite{liu_lead_2013} or Diffnet~\cite{xu_optimal_2019}, as well as some work towards assessing trade off in terms of what network structure will actually provide most reliable estimates with as little computational cost as possible~\cite{yang_optimal_2020, xu_optimal_2019}. To date, there are no rigorous guidelines to prioritise perturbations, but we recommend avoiding difficult perturbations such as those mentioned above involving ring breaking, changes in linker length, changes in charge, and where possible attempt to maximise structural similarity in 2D (via the maximum common substructure) and 3D.

When computing absolute binding free energies, one naturally circumvents the need to generate a perturbation map.

\section{How to analyse benchmark free energy simulations properly}
\label{sec:analysis}

\subsection{Measuring the success of free energy calculations requires careful analysis}

Reliable reporting and analysis of the success of calculations is vital for the validation and benchmarking of free energy methods, as well as the dissemination of published results. The reporting and analysis falls into two major categories -- visualization of the results, and statistical analysis. Here, we make recommendations for both categories.

\subsubsection{Plots of free energy results should adhere to certain common standards}
\label{sec:plotting_results}
\begin{figure*}
    \centering
    \includegraphics[width=0.95\linewidth]{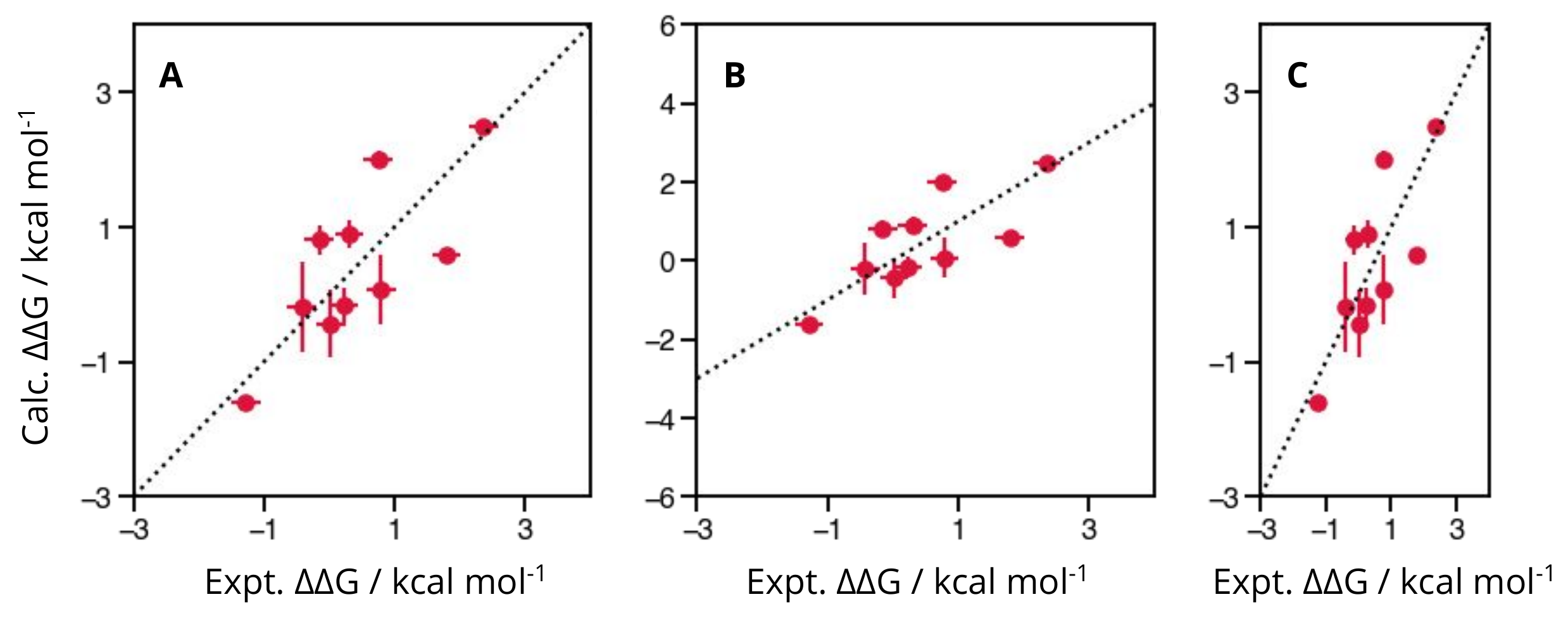}
    \caption{\textbf{Changes to the plotting style can change the appearance of the data.} The above three figures illustrate the same toy data. \textbf{(A)} shows the data correctly, with the same units (which are labelled) and scales on both axes. \textbf{(B)} shows the same data, however the limits on the y-axis have been changed such that the scales is not consistent. \textbf{(C)} is also not consistent, but this is due to the scale of the plot, rather than the limits.}
    \label{fig:plotting-basics}
\end{figure*}

Figures plotting experimental vs. calculated results are a very useful way to gauge the success of a method or a set of calculations. We recommend several key steps to ensure these plots are valuable, communicate accurate information, and are informative and readable. Experimental values (on the x-axis) should be converted into the same units as the free energy results (on the y-axis), and axes should use the same scale. One common issue with plotting free energy results is that different scales are used on the different axes, which can change the appearance of the results, as illustrated in Figure~\ref{fig:plotting-basics}, where changes in the axis and ratios can make the data look more correlated.

Error bars can be very helpful in understanding the uncertainty in the data -- both for calculated and experimental values, and thus both experimental and computational error bars should always be included in visualizations of the data. Different sources of error might be used to quantify this, whether an uncertainty directly from a free energy estimator, variance between repeats or a hysteresis-type analysis. If the experimental errors are not reported, the experimental error can be estimated as e.g. $0.64\,\mathrm{kcal\,mol^{-1}}$ (see Section \ref{sec:exp_uncertainty}). How the error bars have been calculated should be reported in the figure caption.

Additionally, experimental values which were not actually measured (e.g. values resulting from a measured $K_D$ value which only has experimental bounds, such as $> 5\,\mu M$) should not be plotted or should be clearly indicated by different styles and symbols. Such data should not be included in the accuracy or correlation statistics, see  discussion in Section \ref{sec:statistical_analysis}. However, confusion matrices and reporting sensitivity, specificity, and precision can be useful for asserting a models' strength at classifying ligands as binders and non-binders, as demonstrated in~\cite{hauserPredictingResistanceClinical2018}.

Finally, plots of results across multiple targets should typically be shown as one figure per target when the free energy estimates are obtained from the relative free energy calculations. Differences in the success of free energy methods can vary widely between targets, and combining the data across targets onto a single plot can obscure actual performance on any given target. When considering absolute free energies, the affinity ranges between targets may vary, which may result in analysis picking up the correlation between targets and their affinities, rather than the free energy methods ability to differentiate affinities for a particular target. Therefore, if the aim is to evaluate method accuracy per target, each protein-ligand system needs to be studied separately. On the other hand, for absolute free energy calculations it might be of interest to explore whether the method is able to differentiate binding affinities for different targets. One example of such a scenario where considering all target sets together is necessary is free energy calculations for selectivity analysis of similar proteins, where the targets are not independent parameters~\cite{aldeghiPredictionsLigandSelectivity2017}.

\subsubsection{Consistent reporting of statistics, and understanding their limitations is vital for measuring success}
\label{sec:statistical_analysis}
Free energy calculations fall into two categories: absolute and relative. Depending on which type of result are being analyzed --- absolute or relative --- different statistics will be appropriate. Accuracy statistics, such as root mean squared error (RMSE) and mean unsigned errors (MUE) provide information as to how well the computational method recapitulates the experimental results, and allow for a 'best guess' as to how far the computation prediction of new ligands' affinities may be from experiment. Correlation statistics, such as $R^{2}$, Kendall tau ($\tau$) and Spearman's rank ($\rho$) indicate how well a method does at ordering the results, at identifying the best and worst ligand in a set, which in an everyday drug design application, where these models may be used to make purchasing decisions or for synthesis planning, may be a more useful metric than accuracy. However, these statistics can have biases when the number of datapoints (i.e. ligands or edges) are low, as discussed in Section \ref{sec:affinities}.

\begin{figure}
    \includegraphics[width=0.95\linewidth]{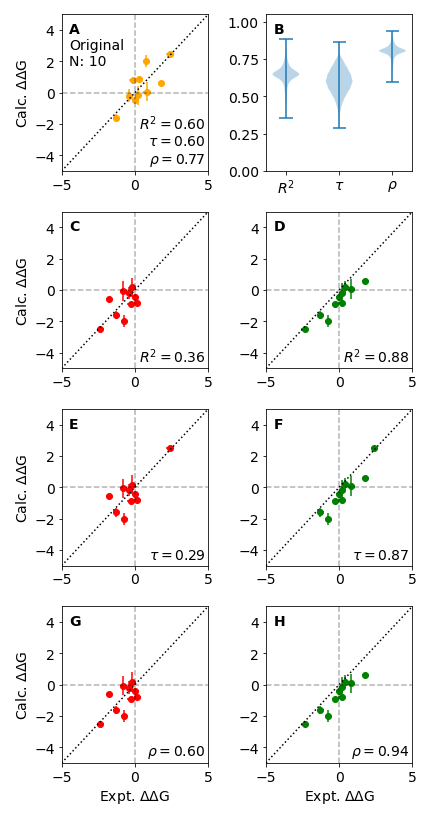}
    \caption{\textbf{Using correlation statistics with relative free energy results are unreliable.} 
    \textbf{(A)} The original set of $N$ datapoints of relative free energy results yields specific statistics for $R^2$, Kendall $\tau$ and $\rho$. However, there are $2^N/2$ possible permutations in the sign for the datapoints, where the changes in sign result in a range of possible statistics from the same underlying data. \textbf{(B)} The distribution of possible values ($2^{10}/2 = 512$) for $R^2$, Kendall $\tau$ and $\rho$ are illustrated in the violin plot. 
    In the following plots (\textbf{(C)}-\textbf{(H)}), the order of permutations are illustrated that result in the lowest (red: \textbf{(C)}, \textbf{(E)} and \textbf{(G)}) and highest (green: plots \textbf{(D)}, \textbf{(F)} and \textbf{(H)}) correlation statistic. 
    The considered statistics are $R^2$ (\textbf{(C)} and \textbf{(D)}), Kendall $\tau$ (\textbf{(E)} and \textbf{(F)}) and $\rho$ (\textbf{(G)} and \textbf{(H)}).
    This illustrates how better correlation statistics for the same relative free energy results can be achieved by simply using different definitions of relative 'directions' for various edges. 
    For this reason, best practise is to avoid reporting correlation statistics for the reporting of relative free energy calculations, and using accuracy  statistics such as RMSE and MUE instead.}
    \label{fig:changing-corr}
\end{figure}

One analysis approach that is commonly a mistake, is the use of correlation-type statistics for the benchmarking of relative free energy calculations without defining a clear protocol for the directions of the perturbations. As relative calculations are pairwise comparisons between ligands, the direction, or sign of the calculation is arbitrary. If a ligand $A$ is $2\,\mathrm{kcal\,mol^{-1}}$ higher affinity than ligand $B$, this could equally be plotted and reported as ligand $B$ being $-2\,\mathrm{kcal\,mol^{-1}}$ lower affinity than ligand $A$. The consequence of the possible inversion of data points can shift the correlation statistics, despite the underlying data being consistent. The same set of data points can give a range of statistical results depending on arbitrary sign-flips in the dataset, where there are $\frac{2^N}{2}$ possible permutations for a set of $N$ relative free energies. While the size of this issue can be affected by the number, range and accuracy of the data points, this can still be problematic, as illustrated in Figure \ref{fig:changing-corr}. If a clear protocol is used, such as mapping all of the results to either be all positive or all negative, or plotting both $A \rightarrow B$ and $B \rightarrow A$ then the statistics quoted will be reproducible, however it is our recommendation to avoid correlation statistics for relative free energy results.

Additionally, correlation statistics, which are appropriate for reporting absolute free energy results, can be sensitive to the number of data points, and the range that they cover, as illustrated in Section \ref{sec:affinities}, Figure ~\ref{fig:N_CI}. This can be exacerbated by experimental uncertainties, which is covered in Section ~\ref{sec:affinities}. Some statistical measures are available that attempt to capture the inherent experimental range in the analysis, such as GRAM ~\cite{cui2019gram} and relative root-mean-squared error (RRMSE). As the number, dynamic range, and experimental uncertainty can all limit the maximum achievable correlation and confidence intervals, it is worth assessing these values \textit{a priori} when deciding if a particular protein-ligand dataset is appropriate for a benchmark (see Section \ref{sec:affinities}).

\subsubsection{Bootstrapping is a reliable method for determining confidence intervals for statistics}

While statistics are a useful measure of the performance of a method, it is also important to understand how accurate those measures are themselves. Is a MUE of $1.2\,\mathrm{kcal\,mol^{-1}}$ much better than $1.3\,\mathrm{kcal\,mol^{-1}}$? Would the performance be likely to change on the addition of new ligands in the series? Is the R$^2$ being heavily leveraged by a handful of outliers? Performing bootstrap analysis allows for confidence intervals to be placed on the statistics, and for these questions to be answered with some confidence. For example, a MUE of $1.2\,(0.6)\,\mathrm{kcal\,mol^{-1}}$ is not statistically different than a MUE of $1.3\,(0.5)\,\mathrm{kcal\,mol^{-1}}$. Bootstrap analysis provides a measure of accuracy to the statistics through random sampling with replacement. Bootstrapping should be performed on the data used to compute the statistic reported --- for relative free energies this illustrate how sensitive the statistics are to the edges chosen, and for absolute free energies: the sensitivity to the ligands in the set. The statistical error for each data point should be incorporated in the bootstrap estimate, where bootstrapping is performed by taking a sample from each data point using its associated variance. It is best practise to report the bootstrapped statistical errors alongside data as 95\% confidence intervals to appropriately evaluate the performance of a particular method, and identify if improvements or changes to a model are statistically significant.

\section{Key learnings}
\label{sec:key_learnings}

\subsection{Analysis tools}
\label{sec:analysis_tools}

We developed a python-based analysis package (\url{http://github.com/openforcefield/openff-arsenic}) to compute statistics from the results of analyzing binding free energy calculations. If statistics from different approaches and sets of calculations are calculated with this package, users can ensure that they are comparing exactly the same statistics calculated in the same way. Results become invariant to different software and definitions of metrics, especially with respect to error or confidence interval calculations. We also see this as a first step towards a containerized benchmarking of methods as is planned for the SAMPL challenges~\cite{mobley_sampl_2021}. There, users will ultimately compare their methods by submissions of containerized methods instead of independently calculated predictions. Thus, all methods will use exactly the same input and their results will be analyzed in the same way. 

For the evaluation of X-ray structure quality, we also provide scripts to calculate Iridium scores and classifications. The Iridium score yields an objective evaluation of the structure.

\subsection{Benchmark set}
\label{sec:benchmark_set}

We assembled a benchmark set using data from prior benchmark studies of relative binding free energy calculations.~\cite{wang_accurate_2015,gapsys_large_2020,schindler_largescale_2020} During evaluations of the given data (Table \ref{tab:struct}), we found quality defects which render parts of the data not appropriate for benchmarking according to our established criteria.

We found deficits in the dataset regarding all our established criteria. There are not trustworthy protein structures (e.g. the PDB 2GMX of Jnk1, Section \ref{sec:struct_data}),
too few data points (e.g. only eight ligands of galectin, \ref{sec:affinities}), or
narrow dynamic ranges (e.g. $0.9\,\mathrm{kcal\,mol^{-1}}$ in BACE\_P2 \ref{sec:affinities}).
We tagged the protein targets as \textit{deprecated} which did not meet a proposed set of minimal criteria (see our Checklist "Minimal requirements for a dataset"). After improvement by addition of new data (such as binding data for additional ligands or binding data spanning a broader dynamic range, availability/use of higher quality protein structures, etc.), these targets could potentially be added to a benchmark set again. 

We acknowledge that the proposed benchmark set does not meet the ideal requirements we established. To date, these requirements are challenging to meet due to scarce high-quality experimental data, especially after applying all the criteria we lay out. Experimental affinity measurements from a single source often do not cover dynamic ranges $>5\,\mathrm{kcal\,mol^{-1}}$ and much larger dynamic ranges become unrealistic. Large numbers of single source affinity data points are rarely available and additionally impose the practical limitations of large computational resources for benchmarking calculations. As a chain is only as strong as its weakest link, the above points need to be paired with high-quality structural data and a good preparation of the simulation input. 

We welcome community contributions and assistance to build a benchmarking dataset that will eventually fulfill our high standards.

\section{Recommendations}
\label{sec:recommendations}

Methods for binding free energy calculations have been continuously developed over the last decades and are increasingly used both in academic research, as well as pharmaceutical industry applications in structure based drug discovery~\cite{sherborne_collaborating_2016, gapsysLargeScaleRelative2020,schindler_largescale_2020}, making their validation and benchmarking particularly crucial. 

In order to reliably benchmark methods, we provide best practices recommendations for setting up benchmark calculations.
This setup begins with the appropriate choice of experimental inputs and data, which includes the choice of target(s) and ligands (Section~\ref{sec:dataset}). 
We require both
structural information (Section \ref{sec:struct_data}) and 
affinity data (Section~\ref{sec:affinities}.
This input information needs to be adequately prepared to generate simulation inputs (Section \ref{sec:prep}) before the systems are simulated with a specific choice of 
software, calculation setup, and simulation protocol. 
Here, we made a variety of recommendations as to how to select and prepare systems for benchmarking.

Benchmarking also requires analysis and comparison with experiment, thus we also recommend standard reporting procedures (Section~\ref{sec:analysis}). These provide a mechanism to assess the accuracy of the calculations, present the results and compare to calculations done with other methods. These standard procedures will make it far easier to compare results across studies done by different researchers or using different tools.

Our recommendations are exemplified in publicly available tools for the analysis of calculations (\url{http://github.com/openforcefield/openff-arsenic})
and a living protein-ligand benchmark dataset (\url{http://github.com/openforcefield/protein-ligand-benchmark}).
This set is living in the sense that we expect it to be subject to ongoing updates, curation, and improvement -- both by ourselves and by the community, and we welcome community input via the GitHub issue tracker at \url{http://github.com/openforcefield/protein-ligand-benchmark/issues}. Additionally, further curation is clearly necessary as our recommendations are in part not fulfilled in the initial version of this benchmark dataset. Partially, this is because we have begun from previously used benchmark sets and are beginning the curation process, but also because it is difficult to find large and accurate experimental datasets meeting all the desired characteristics. Thus, in our initial set, the relevant issues are annotated and we expect the benchmark set to evolve to better meet the recommendations given here.

We hope that our recommended best practices will be adopted and where necessary improved by the community. We believe that these best practices will ultimately help advance the accuracy, applicability, and availability of binding free energy calculations.

\section{Checklists}
\label{sec:checklists}
\begin{Checklists*}[hp!]
\begin{checklist}{Choose Suitable Protein Structures for Benchmarking}
\textbf{Find experimental structural data: Section~\ref{sec:struct_data}}

    \textbf{Global criteria}
    \begin{itemize}
        \item Select the best available structure using DPI or coordinate error (< 0.7)
        \item Ensure experimental data is available, i.e. electron/neutron density or cryo EM map
        \item Ensure the reported $Rfree < 0.45$ when resolution $\le 3.5~$\AA{}
        \item Ensure that the reported difference between R and $Rfree \le 0.05$
        \end{itemize}
    \textbf{Local criteria}
    \begin{itemize}
        \item Determine if there are crystal contacts and assess if they affect protein conformation. Select structures with no crystal packing atoms within $6\,$\AA{} of any ligand atom.
        \item Confirm that the ligand has at least partial density (check visually or real space correlation coefficient (RSCC) > 0.90) and the density is adequate to confirm ligand presence and binding mode
        \item Ensure that all ligand and active site atoms have occupancy >0.80
        \item Identify active site atoms with partial density and confirm these are acceptable and not key contacts
        \item Confirm active site crystallographic waters have density and no difference density
        \item Identify any alternate conformations for ligand and active site atoms. Select the alternate conformation with the highest occupancy and fewest clashes. 
        \item Confirm that the ligand is not covalently bound as deposited, and is also not likely to have reacted to become one
        \item Check for any missing loops or residues and side chain atoms in the structure and confirm these are not near the binding site/not key for the study
    \end{itemize}
\end{checklist}

\begin{checklist}{Affinity Data}
\textbf{Find experimental affinity data: Section~\ref{sec:affinities}}
    \begin{itemize}
    \item Select single source data.
    \item Extract binding data from original source and convert carefully.
    \item Remove data points outside detection limits.
    \item Ideally data should be from biophysical assays. With functional assays, more care must be taken.
    \item Assess dataset quality in terms of number of datapoints, experimental affinity range and experimental error to know the maximally achievable precision.
    \end{itemize}
\end{checklist}
\end{Checklists*}
\begin{Checklists*}[hp!]
\begin{checklist}{Prepare the System with Care Because Failures Here are Crucial}
\textbf{Prepare structural data for simulation: Section~\ref{sec:prep}}
    \begin{itemize}
        \item Assess which domains of the X-ray structure are needed and retain domains present in the experimental study, unless it is known that further simplifications can be made without affecting accuracy.
        \item Check other components (cofactors, crystallographic waters, other ligands, PTMs) of the structure and make sure you include everything which is key for the study.
        \item Split the protein and ligand structures to prepare separately.\\
        \textbf{Protein preparation}
    \begin{itemize}
                \item Add caps if the structure's termini are not resolved.
                \item If possible, model missing loops, if loops are too long (> 15 to 20 residues) or too mobile, consider capping the ends and adding restraints, or modeling a short glycine loop that links both ends. These must not be in the binding site.
                \item Inspect for side chain flips of side chains which can fit density similarly when reoriented (HIS, ASN, GLN); confirm that the orientations chosen lead to preferred interactions with the ligand. Evaluate alternate placements if necessary.
                \item Check the protonation states of the ligand and receptor, again checking in the context of the interactions that would be formed with the ligand.
    \end{itemize}
        \textbf{Ligand preparation}
    \begin{itemize}
            \item Ensure that the chemical structure is correct (bond orders, stereochemistry).
            \item Align the ligand series based on conformations of (X-ray) reference compound(s).
            \item Check tautomer and ionization states. Determine whether multiple possibilities need to be considered.
            \item Check whether alternate rotamers may need to be considered after alignment to reference compound(s).
    \end{itemize}
    \end{itemize}
        \textbf{System preparation}
    \begin{itemize}
        \item Assemble the protein, ligand and cofactors. 
        \item Without removing crystallographic waters and ions, solvate the complex or embed it in a membrane. 
        \item Add ions; use an appropriate salt concentration (sodium and chloride ions) to model the assay.
        \item Equilibrate the system.
    \end{itemize}
\end{checklist}

\begin{checklist}{Carefully Select Appropriate Simulation Details}
\textbf{Choose simulation setup: Section~\ref{sec:alchemical_prep}}
    \begin{itemize}
        \item Choose absolute vs. relative calculations.
        \item Choose topology approach and alchemical pathway.
        \item Choose sampling protocol.
        \item Plan a perturbation map if calculations are relative.
    \end{itemize}
\end{checklist}
\end{Checklists*}
\begin{Checklists*}[hp!]
\begin{checklist}{Present Graphs of Results in a Consistent Manner}
\textbf{Presenting results in an appropriate format: Section~\ref{sec:plotting_results}}
\begin{itemize}
\item Clearly label the data with titles, legends, and captions.
\item Plot results with the dependent variable (calculated) on y-axis, and the independent variable (experimental) on the x-axis. 
\item Ensure that the data are reported in the same units on both axes, and labelled. The scale of the axis in real space should be consistent, such that a 1 cm change on the x-axis corresponds to the same change in affinity to 1 cm on the y-axis.
\item Plot only one target per plot, unless specifically looking at selectivity or considering multiple systems by means of absolute free energy calculations.
\end{itemize}
\end{checklist}

\begin{checklist}{Use Careful Statistical Analysis to Quantify Performance}
\textbf{Quantifying the success of a method: Section~\ref{sec:statistical_analysis}}
\begin{itemize}
\item Identify which metrics are appropriate for your method. Statistics that measure accuracy, such as RMSE and MUE, are commonplace; correlation statistics are appropriate for absolute free energies, and relative free energies only when perturbation map and $\Delta G$ direction is consistent among benchmarked methods.
\item Bootstrap statistics to provide confidence intervals. 
\item Provide confidence intervals for all reported values and avoid overinterpreting results given these intervals.
\end{itemize}
\end{checklist}

\begin{checklist}{Minimal and Ideal Requirements for a Dataset}
\textbf{Summary of the most important points from the checklists above and definition of minimal as well as ideal (boldface) requirements for a benchmark set.}
    \begin{itemize}
        \item Experimental structure should be Iridium classified as at least MT (\textbf{ideally HT}).
        \item Single source experimental activities (\textbf{ideally from a biophysical assay}).
        \item At least \ligcountthres{} (\textbf{ideally 25}) data points/ligands.
        \item A dynamic range of at least $>\ligrangethres{}\,\mathrm{kcal\,mol^{-1}}$ ( \textbf{ideally $>5\,\mathrm{kcal\,mol^{-1}}$}).
        \item Well prepared structures (charge and tautomeric states) checked by at least one other experienced person (\textbf{ideally by the community}).
    \end{itemize}
\end{checklist}
\end{Checklists*}

\pagebreak
\section{Author Contributions}
%
  \textbf{CIB} contributed to the text in the Section \ref{sec:prep} \nameref{sec:prep}.\\
  \textbf{HEBM} wrote Sections \ref{sec:analysis} \nameref{sec:analysis} and \ref{sec:analysis_tools} \nameref{sec:analysis_tools}, created the figures therein, and contributed to Section \ref{sec:affinities} \nameref{sec:affinities}.\\
  \textbf{JDC} helped in the manuscript design and edited the final manuscript.\\
  \textbf{DFH} coordinated the manuscript preparation, wrote the Section \ref{sec:struct_data}  \nameref{sec:struct_data} and \ref{sec:recommendations} \nameref{sec:recommendations}, contributed to most other sections.\\
  \textbf{ASJSM} wrote Section~\ref{sec:alchemical_prep} \nameref{sec:alchemical_prep} and created Figure~\ref{fig:map}, Figure~\ref{fig:protocols}, updated Figure~\ref{fig:system_preparation} and helped edit the manuscript. \\
  \textbf{DLM} helped outline some of the sections, edited many of the sections and contributed to the text in several sections. \\
  \textbf{LPB} wrote Section~\ref{sec:prep} \nameref{sec:prep}.\\
  \textbf{CS} wrote Section~\ref{sec:affinities} \nameref{sec:affinities} and created Figure~\ref{fig:N_CI}.\\
  \textbf{GT} wrote Sections~\ref{sec:intro} \nameref{sec:intro} and \ref{sec:dataset} \nameref{sec:dataset}, edited the final manuscript.\\
  \textbf{GLW} generated the data for Table \ref{tab:struct} and contributed to the text in the Section \ref{sec:struct_data}  \nameref{sec:struct_data}.\\
  \textbf{VG} contributed to several sections mainly regarding the absolute binding free energy calculations.\\
For a more detailed description of author contributions,
see the GitHub issue tracking and changelog at \githubrepository.

\section{Other Contributions}
%

We want to thank the authors of the following publications for establishing the initial protein-ligand benchmark sets: Wang et al.\cite{wang_accurate_2015}, Perez-Benito et al.\cite{perez-benito_predicting_2019}, Gapsys et al.\cite{gapsys_large_2020} and Schindler et al.\cite{schindler_largescale_2020}.

For a more detailed description of contributions from the community and others, see the GitHub issue tracking and changelog at \githubrepository.

\section{Potentially Conflicting Interests}

DLM serves on the scientific advisory board for OpenEye Scientific Software and is an Open Science Fellow with Silicon Therapeutics.
ASJSM is a consultant for Exscientia.
JDC is a current member of the Scientific Advisory Board of OpenEye Scientific Software and a consultant to Foresite Laboratories.
HEBM is employed by MSD.
\section{Funding Information}
JDC acknowledges support from NIH grant P30 CA008748.
We appreciate the financial support of the Open Force Field Consortium (\url{https://openforcefield.org}), and the National Institutes of Health (NIGMS R01GM132386 and R01GM108889).
HEBM acknowledges support from a Molecular Sciences Software Institute Investment Fellowship and Relay Therapeutics.

\section*{Author Information}
\makeorcid

\bibliography{livecoms-sample}



\end{document}